\documentclass[usenatbib,usegraphicx,twocolumn,a4paper]{mn2e}
\usepackage[latin1]{inputenc}
\usepackage{graphicx}
\usepackage{amsmath}
\usepackage{times}
\usepackage{amssymb}
\usepackage{natbib}
\usepackage{rotating}
\usepackage{lscape}
\usepackage{multirow}
\usepackage{multirow}
\usepackage{csquotes}
\usepackage[usenames,dvipsnames]{xcolor}

\newcommand{\vb}{\textcolor{cyan}}
\renewcommand{\vb}{\textcolor{black}}
\newcommand{\psc}{\textcolor{black}}

\newcommand{\oldversion}[1]{}
\title[Probing DE models with EVS of galaxy clusters from the DEUS-FUR simulations]{Probing dark energy models with extreme pairwise velocities of galaxy clusters from the DEUS-FUR simulations}
\author[V.~R.~Bouillot et al.] {Vincent~R.~Bouillot,$^{1,2}$\thanks{vincent.bouillot@uct.ac.za} Jean-Michel~Alimi,$^{2,3}$\thanks{jean-michel.alimi@obspm.fr} Pier-Stefano Corasaniti $^2$\thanks{pier-stefano.corasaniti@obspm.fr} and \newauthor Yann~Rasera$^{2}$\thanks{yann.rasera@obspm.fr} \\ $^1$ Centre for Astrophysics, Cosmology \& Gravitation, Department of Mathematics \& Applied Mathematics, \\  University of Cape Town, Cape Town 7701, South Africa \\ $^2$ CNRS Laboratoire Univers et Th\'eories (LUTh), UMR 8102 CNRS, Observatoire de Paris, \\ Universit\'e Paris Diderot,  5 Place Jules Janssen, 92190 Meudon, France \\  $^3$ Institut d{'}Astrophysique de Paris, UMR 7095 CNRS, Universit\'e Pierre et Marie Curie, 98bis Blvd Arago, 75014 Paris, France}

\begin{document}
\date{\today}
\maketitle

\begin{abstract}
Observations of colliding galaxy clusters with high relative velocity probe the tail of the halo pairwise velocity distribution with the potential of providing a powerful test of cosmology. As an example it has been argued that the discovery of the Bullet Cluster challenges standard $\Lambda$CDM model predictions. Halo catalogs from N-body simulations have been used to estimate the probability of Bullet-like clusters. However, due to simulation volume effects previous studies had to rely on a Gaussian extrapolation of the pairwise velocity distribution to high velocities. Here, we perform a detail analysis using the halo catalogs from the Dark Energy Universe Simulation Full Universe Runs (DEUS-FUR), which enables us to resolve the high-velocity tail of the distribution and study its dependence on the halo mass definition, redshift and cosmology. Building upon these results we estimate the probability of Bullet-like systems in the framework of Extreme Value Statistics. We show that the tail of extreme pairwise velocities significantly deviates from that of a Gaussian, moreover it carries an imprint of the underlying cosmology. We find the Bullet Cluster probability to be two orders of magnitude larger than previous estimates, thus easing the tension with the $\Lambda$CDM model. Finally, the comparison of the inferred probabilities for the different DEUS-FUR cosmologies suggests that observations of extreme interacting clusters can provide constraints on dark energy models complementary to standard cosmological tests.
\end{abstract}


\section{Introduction}
The advent of large scale cluster survey programs \citep[][]{PLANCK,SPT,XXL,ACT} will soon provide large catalogs of galaxy clusters to test the standard cosmological scenario based on the Cold Dark Matter paradigm with Cosmological Constant ($\Lambda$CDM). 

Galaxy clusters are the largest observable structures in the universe which reside inside massive Dark Matter (DM) halos. These are gravitationally bound objects that result from the gravitational collapse of small matter density fluctuations which were present in the early universe. Because of this the number density of DM halos carries complementary information on both the statistics of the primordial matter density field and the growth rate of structures. Measurements of the abundance of cluster of galaxies aim to probe such features, but their effectiveness to constrain cosmological models depends upon the precise understanding of the survey selection function \citep{P2006,XXL,C2012} as well as the availability of accurate measurements of cluster masses \citep[see e.g.][]{MM2004,LH2004,C2009}. However, measuring the mass of several hundreds of clusters is a challenging task that requires costly follow-up observations of individual objects, thus inevitably spanning these projects over long periods of time. While completing these programs, cosmological information can still be inferred from reduced datasets consisting of the most massive objects. These probe the high-mass end of the halo mass function and thus have the potential to rule out entire classes of cosmological models. The advantage is that such systems, being also the most luminous, are easier to detect and being limited in number makes their follow-up observations more readily accessible.

In recent years this complementary approach to cluster cosmology has received lots of attention due to the discovery of very massive clusters at high-redshift \citep[][]{J2009,R2009,F2011,Menanteau2012a,S2012}. These detections have lead several authors to question the basic assumptions of the concordance $\Lambda$CDM model \citep[][]{JimenezVerde2009,Hoyle2011,HP2012}. However, estimating the probability that such extreme clusters occur is far from being trivial. As pointed out by \cite{Hotchkiss2011} assessing the rareness of clusters in terms of the probability of observing at least one cluster of mass larger than that observed and/or at a higher redshift can lead to biased conclusions. 

A natural framework to address these questions is given by the Extreme Value Statistics (EVS). This can be used to predict the probability distribution of the mass of the most massive halo in the sample from prior knowledge of the halo mass function in a given cosmological model. This has been the subject of several studies \citep[see e.g.][]{Davis2011,Waizmann2011,HarrisonColes2011} which have focused on the mass as measure of cluster extremeness. However, a careful analysis of the most massive systems so far observed suggests that there are other characteristics that are indicative of the extremeness of these objects. For instance, 1E0657-56 \citep{Markevitch2002a,Markevitch2006a}, MACS J0025.4-1222 \citep{Bradac2008a}, ACT-CL \mbox{J0102-4915} \citep{Menanteau2012a} and AS1063 \citep{Gomez2012a} are merging clusters with high relative velocities. Among these 1E0657-56 is one of the most well documented. Known as the \enquote{Bullet Cluster}, it is composed of two clusters which have undergone a nearly head on collision. The main cluster has a mass $\simeq 10^{15}$ h$^{-1}$ M$_\odot$, while the smaller one has mass $\simeq 10^{14}$ h$^{-1}$ M$_\odot$ \citep{Clowe2004a}. The system is located at $z=0.296$ \citep{Clowe2006,Bradac2006} and the clusters are separated by a distance of \mbox{$\simeq 0.51$ h$^{-1}$ Mpc}. X-ray observations have shown that the collision has ripped off the clusters of their intra-cluster gas which is trapped in a shock with Mach number close to $\sim 3$. This implies a velocity of the shock front of \mbox{$\sim 4700$ km s$^{-1}$} \citep{Markevitch2006a}, which has been usually interpreted as being also the relative velocity of the colliding clusters. Under this hypothesis \citet{Hayashi2006a} analysed the velocities of the sub-halo distribution from the Millenium Simulation \citep{Springel2005a} and concluded that the existence of the Bullet Cluster is consistent with the standard $\Lambda$CDM cosmology. However, due to the limited volume of Millenium Simulation their result do not rely on direct measurement but rather on extrapolating the sub-halo velocity probability distribution to host halos with mass $>10^{14}$ h$^{-1}$ M$_\odot$. Furthermore, the relative velocity of the colliding clusters may well be different from that of the gas. As shown by \citet{Milosavljevic2007a} using 2-D hydrodynamical simulations this can be up to $\simeq 15 \%$ smaller \citep[see also][]{SpringelFarrar2007}. 

To date the most detailed study of the Bullet Cluster has been performed by \citet{Mastropietro2008a} who have used 3-D non-cosmological hydrodynamical simulations to determine the initial physical configurations of the colliding clusters resulting in a system whose properties reproduce those observed in the Bullet Cluster. These include the displacement between the X-ray peaks and the mass distribution, the morphology of the shock velocity, the surface brightness and the projected temperature profiles across the shock. \citet{Mastropietro2008a} have shown that the colliding halos must have an initial velocity \mbox{v$_{12} \sim 3000$ km s$^{-1}$} with a mass ratio of $6:1$, an initial separation of $5$ Mpc (implying an initial redshift of $z=0.486$) and a collision angle $\theta_{12}$ close to 0 (i.e. head on collision). The identification of these parameters has provided criteria to select Bullet-like halo pairs in cosmological simulations, a crucial step to estimate the probability of finding the Bullet Cluster in a given cosmological setup. 

\citet{Lee2010a} analysed the halo catalog from the MICE simulations \citep{Crocce2010} of $\Lambda$CDM cosmology at $z=0$ and $0.5$ to infer the pairwise velocities probability distribution for different mass ratios, distance separation and relative velocity. Quite remarkably they found that none of the analysed catalogues contains a system with parameters corresponding to that of the Bullet Cluster. Nevertheless, by fitting the probability density distribution to a Gaussian, they were able to extrapolate the probability to high relative velocities. They found the rate of occurrence of Bullet Cluster-like systems to be $P({\rm v}_{12}>3000\,\text{km s}^{-1})=3.3\times10^{-11}$ and $3.3\times10^{-9}$ at $z=0$ and $z=0.5$ respectively. \citet{Thompson2012a} performed a similar analysis of a set of $\Lambda$CDM simulations with DM mass resolution varying from 9$\times 10^9$ to 5.7$\times 10^{11}$ h$^{-1}$ M$_\odot$ and box sizes ranging from $200$ h$^{-1}$ Mpc to $2$ h$^{-1}$ Gpc. By extrapolating the cumulative distribution to high relative velocities these authors obtained $P({\rm v}_{12}>3000\,\text{km s}^{-1})=2.76\times10^{-8}$ at $z=0.489$ for masses $\ge 10^{14}$ M$_\odot$. \psc{These probabilities imply that in the observable cosmic volume the existence of the Bullet Cluster pair is either incompatible with the $\Lambda$CDM scenario} or as argued by \citet{Thompson2012a} that the initial conditions determined from the analysis of non-cosmological hydrodynamical simulations by \citet{Mastropietro2008a} have to be revised to much lower values of v$_{12}$ \cite[see e.g.][for a recent study]{Lage2013a}. 

A critical point is that all these analyses rely on extrapolating the tail of the pairwise velocity probability distribution to high-velocities. This is a direct consequence of the limited volumes of the numerical simulations from which these results have been derived. Furthermore, the probability of observing the Bullet Cluster has been directly estimated from the tail of the probability density distribution fitted to a Gaussian \psc{which may suffer of potential biases especially if the tail of the distribution is non-Gaussian.}

In the work presented here we improve these studies in several ways. We use the catalog of halos from the Dark Energy Universe Simulation - Full Universe Runs (DEUS-FUR) which cover the entire observable cosmic volume with a box-length of $21$h$^{-1}$ Gpc and $8192^3$ dark matter particles \citep{Alimi2012,Rasera2014}. These simulations provide an unprecedented large statistical sample to test the rareness of halo properties for different cosmological models. The large simulation volume allows us to perform a detailed analysis of the pairwise velocity especially in the high-velocity tail. Building upon this study we use the Extreme Value Statistics (EVS) to quantify the probability of observing the Bullet Cluster. This enables us for the first time to perform a cosmological model comparison of the Bullet Cluster observation.

The paper is organized as follows. In Section \ref{Sec2}, we introduce the N-body simulation data, the halo finder algorithm and the halo pair selection method. In Section \ref{dndv_fof}, we described the dependence of the pairwise velocity function on the halo finder parameter and discuss the physical implications, while in Section \ref{sec_cosmoz} we study the redshift evolution and cosmology dependence. In Section \ref{evsbull} we introduce the Extreme Value Statistics and present the results of its application to the Bullet Cluster. Finally in Section \ref{conclu} we discuss our conclusions.

\section{N-Body Simulation Dataset}
\label{Sec2}
\subsection{DEUS Full Universe Runs}

We use numerical data issued from the DEUS-FUR project. This consists of three N-body simulations with $8192^3$ dark matter particles and box size of \mbox{(21000 h$^{-1}$ Mpc)$^3$} enclosing the observable volume of a flat $\Lambda$CDM cosmology and two dark energy models with different expansion histories. The simulations have been realized using the application AMADEUS - A Multi-purpose Application for Dark Energy Universe Simulation - expressly developed for the DEUS-FUR project \citep{Alimi2012}. This includes the generator of Gaussian initial conditions for which we use an optimized version of the code MPGRAFIC \citep{prunet08}, the N-body solver which is a version of the RAMSES code \citep{teyssier02} specifically improved to run on a large number of cores (\mbox{$>$ 40000} MPI tasks in production mode) and an optimized Friend-of-Friend halo finder \citep{Roy2014}. RAMSES solves the Vlasov-Poisson equations using an Adaptive Mesh Refinement (AMR) Particle Mesh method with the Poisson equation solved with a multigrid technique \citep{guillet11}. A detailed description of the algorithms, optimization schemes and the computing challenges involved with the realization of DEUS-FUR is given in \cite{Alimi2012}. All simulations share the same phase of the initial conditions. The coarse grid of the AMR hierarchy contains $8192^3$ resolution elements and is allowed to refined six times, reaching a formal spatial resolution of 40 h$^{-1}$ kpc, while the particle mass resolution for the different models is \mbox{$m_p \simeq 10^{12}$ h$^{-1}$ M$_\odot$}. Such resolution roughly corresponds to the size and mass of the Milky Way.

\begin{figure*}
\begin{minipage}{17cm}
\begin{center}
\begin{tabular}{cc}
  \includegraphics[width=8cm,height=6cm]{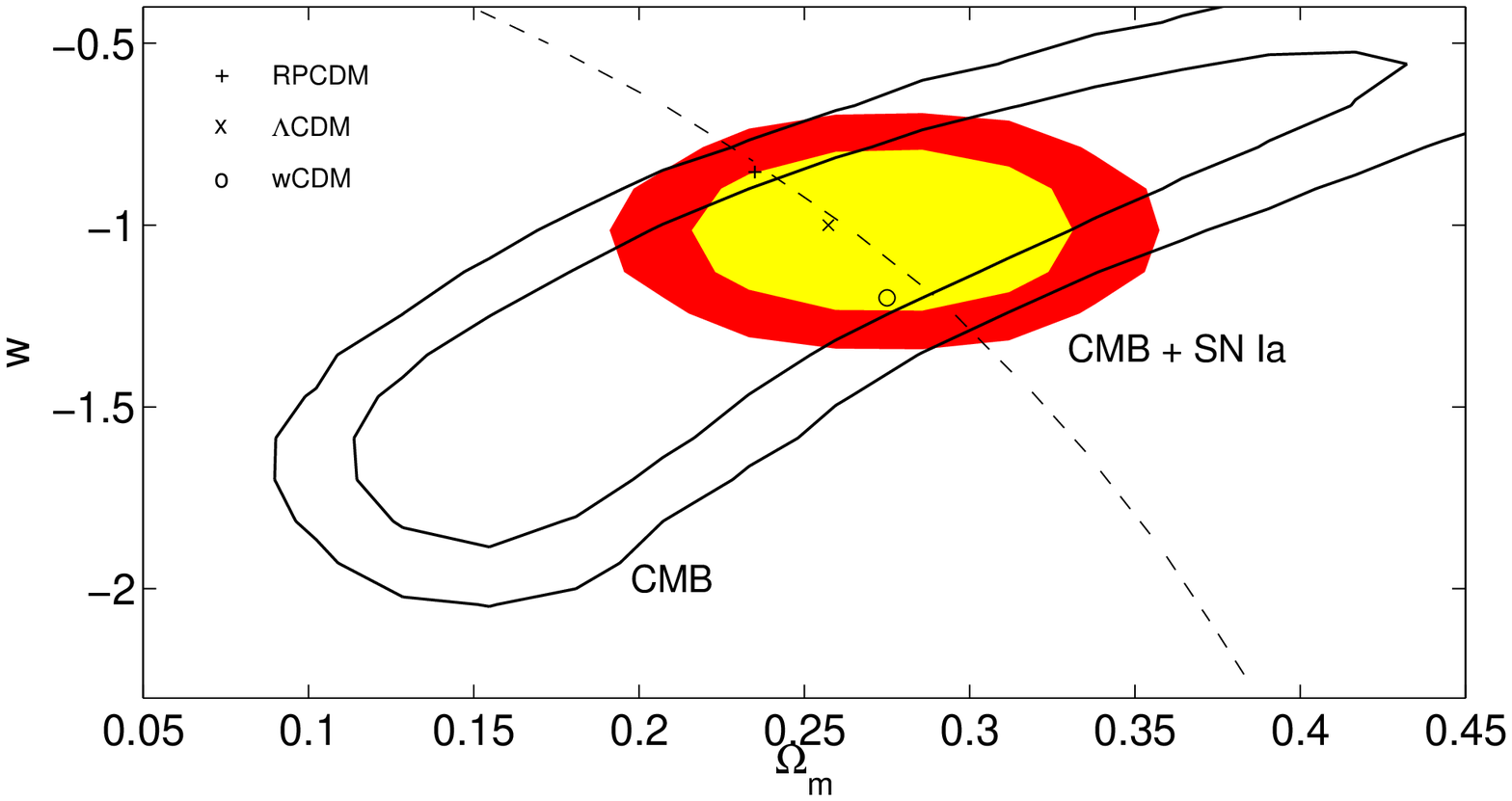}&
  \includegraphics[width=8cm,height=6cm]{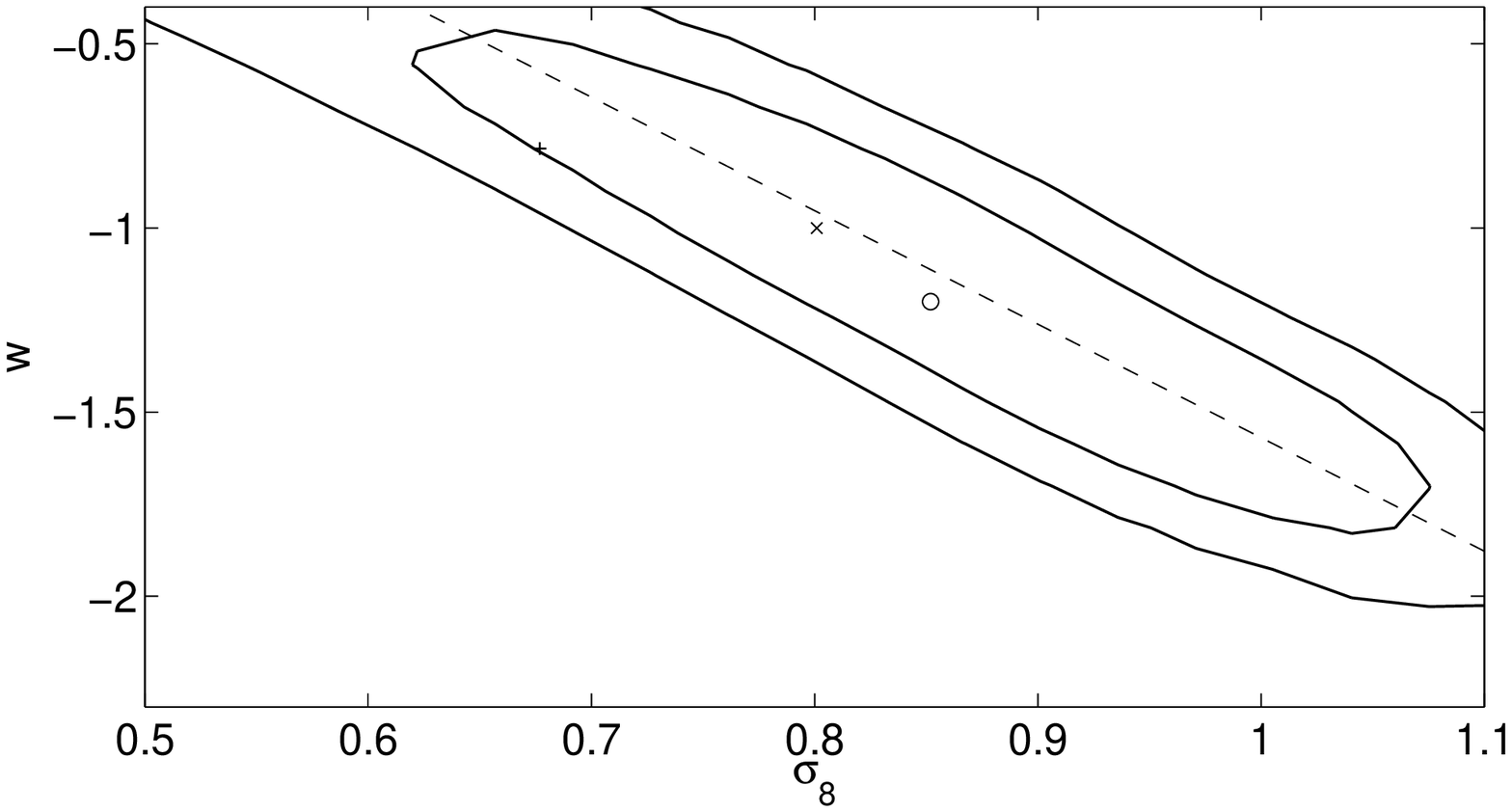}
\end{tabular}
  \caption{\label{contours}Left panel: Marginalized $1$ and $2\sigma$ contour plots in the $\Omega_m-w$ planes from the likelihood analysis of the CMB data from WMAP-7yr observations (solid lines) and in combination with luminosity distance measurements to SN Ia from the UNION dataset (filled contours). The dashed line indicates the luminosity distance degeneracy curve. Right panel: Marginalized $1$ and $2\sigma$ contour plots in the $\sigma_8-w$ plane from CMB data only (solid lines). The dashed line indicates the degeneracy curve from the CMB anisotropy power spectra.}
\end{center}
\end{minipage}
\end{figure*}

The simulated cosmologies consist of a flat $\Lambda$CDM model best-fit to the WMAP-7yr data \citep[$\Lambda$CDM-W7,][]{spergel07}, a quintessence model with Ratra-Peebles potential \citep[RPCDM,][]{RatraPeebles} and a ``phantom'' dark energy model with constant equation of state $w<-1$ \citep[$w$CDM,][]{KamionCald}. 

The model parameters have been calibrated to fit the Cosmic Microwave Background anisotropy power spectra from WMAP 7-year observations \citep{spergel07} and luminosity distances measurements to Supernova Type Ia from the UNION dataset \citep{union}. In particular, the mean cosmic matter density, $\Omega_m$, has been chosen within the marginalized $\sim 1\sigma$ contour in the $\Omega_m-w$ plane along the degeneracy line of the SN Ia data (see left panel in Fig. \ref{contours}); while the values of the root-mean-square fluctuation amplitude of the density contrast at $8$ h$^{-1}$ Mpc, $\sigma_8$, has been chosen within the marginalized $\sim1\sigma$ confidence contours in the $\sigma_8-w$ plane nearly parallel to the degeneracy line of the CMB data (see right panel in Fig. \ref{contours}). This particular choice is motivated by the fact that through the analysis of the DEUS-FUR simulations we aim to test whether observables of the non-linear clustering of matter can break the degeneracies affecting current cosmological parameter constraints. A summary of the cosmological model parameter values and the simulation characteristics are reported in Table \ref{DEUSSALL1}.

\setcounter{table}{0}
\begin{table}
\begin{center}
\begin{tabular}{cccc}
\hline
Parameters & RPCDM & $\Lambda$CDM-W7 & $w$CDM \\
\hline
\hline
$\Omega_m$  & 0.23& 0.2573 & 0.275 \\
$\Omega_b h^2$  & 0.02273& 0.02258 & 0.02258 \\
$\sigma_8$& 0.66 & 0.8  & 0.852 \\
$w_0$& -0.87 & -1  & -1.2 \\
$w_1$ & 0.08 & 0 & 0 \\
\hline
$z_{ini}$  & 94 & 106 & 107 \\
$m_p ($h$^{-1}$ M$_\odot$)  & $1.08 \times 10^{12}$ & $1.20 \times 10^{12}$  & $1.29 \times 10^{12}$ \\
$\Delta x$ (h$^{-1}$ kpc) & 40 & 40 & 40 \\
\hline
\end{tabular}
\caption{Cosmological parameter values of the DEUS-FUR simulated cosmologies. For all models the scalar spectral index is set to $n_s=0.963$ and the Hubble parameter $h=0.72$. We also report the values of a linear equation of state parameterization $w(a) = w_0 + w_1(1- a)$ for the different models (see \citet{Alimi2010} for details). In the bottom table we list the values of the initial redshift of the simulations $z_{ini}$, the particle mass $m_p$ and the comoving spatial resolution  $\Delta x$. For all three simulations the box-length is L$_{\textrm {box}}=21000$~h$^{-1}$ Mpc and the number of dark matter particles is $8192^3$.}
\label{DEUSSALL1}
\end{center}
\end{table}

\subsection{Halo Finder and Halo Pair Selection}
The detection of halos in the DEUS-FUR simulations is performed with a highly-scalable parallelized version of the Friend-of-Friend halo finder algorithm \citep[][]{Roy2014} implemented in the AMADEUS application. This algorithm detects halos as groups of particles characterized by an interparticle distance smaller than a given linking length (in units of the mean interparticle distance) or percolation parameter $b$. 

In the study presented in Section \ref{dndv_fof} we consider halos detected with $b=0.1,0.15$ and $0.2$ respectively. In order to limit the effect of numerical artifacts due to low number of particles we only consider halos with $>100$ particles, which corresponds to halo masses \mbox{{$M\gtrsim 10^{14}$h$^{-1}$ M$_\odot$.}}

\begin{table}
\begin{center}
\begin{tabular}{ccccc}
\hline
&Redshift& RPCDM & $\Lambda$CDM-W7  & $w$CDM \\
\hline
\hline
\multirow{3}{*}{\rotatebox{90}{Single}} & $0$ & 76,180,615 & 144,630,773 & 169,186,215\\
& $0.3$& 40,613,387 & 90,788,115  & 109,227,390\\
&$0.5$& 24,554,151 & 61,804,451  & 74,966,075 \\
\hline
	\hline
	\multirow{3}{*}{\rotatebox{90}{Pair}} & $0$ & 47,727,489 & 125,555,136 & 156,600,237 \\
& $0.3$& 17,297,267 & 58,501,507  & 76,027,601 \\
&$0.5$& 7,700,836 & 31,192,185  & 40,877,856 \\
\hline
\end{tabular}
\caption{Total number of FoF(b=0.2) single halos with $>100$ particles and pair of halos with separation $<15$ h$^{-1}$ Mpc detected in the comoving volume of the three cosmological DEUS-FUR simulations at redshift $z=0,0.3$ and $0.5$ respectively.}
\label{NZ}
\end{center}
\end{table}

In Table \ref{NZ} we report the total number of FoF(b=0.2) halos detected in the DEUS-FUR simulations at different redshifts. These are vast halo catalogues for which the selection of halo pairs and the calculation of relevant quantities poses a challenging computational problem. In fact, the complexity of a standard brute force computation of the relative velocities for all pair separations grows with the square of the number of halos. This leads to a prohibitive computational time as soon as the number of halos exceeds $\sim 10^6$. However, we are interested only in halo pairs characterized by a small distance separation such as the Bullet Cluster initial configuration found in \citet{Mastropietro2008a}. Therefore, we can significantly reduce the number of computations by using an octree space decomposition which enables the computation of pairs up to a maximum distance $d_{12}^{\rm max}$. To be conservative we set $d_{12}^{\rm max}\approx3\,R_{\rm max}$, where $R_{\rm max}$ is the radius of the most massive halo in the catalogues. This gives a maximal distance separation \mbox{$d_{12}^{\rm max}\sim 15$ h$^{-1}$ Mpc}. Thanks to such a space decomposition we compute all relevant quantities for all pairs within the pruning radius, thus avoiding the most time consuming long-range computations. The implementation of this algorithm allows us to compute the relative velocities of about 100 million pairs in less than 2 minutes on a 64 core (Intel Xenon CPU X7550$@$2.00Ghz) local machine. The number of pairs within 15 h$^{-1}$ Mpc for the three cosmological catalogs at three redshifts of interest is given in Tabel \ref{NZ}. As we can see at $z=0$ the number of pairs is a significant fraction of the total number of halos detected in the simulations which is indicative of the high level of clustering of such objects compared to that of the average density field with a mean inter-halo separation of 40-70 h$^{-1}$ Mpc. 

\section{Pairwise Velocities and Friend-of-Friend Halos}\label{dndv_fof}
In this section we study the dependence of the pairwise velocity function on the halo definition specified by the value of the percolation parameter $b$. It is usually understood that for a given value of $b$ the FoF algorithm selects halos whose boundary has approximatively a fixed isodensity surface. For instance, the local surface overdensity with respect to the mean matter density of two particles within a sphere of radius $b$ is $\delta_{\textrm{FOF}}\sim 1/ 2b^3$ \citep{Summers1995,Audit1998}. In the case $b=0.2$ this gives $\delta_{\textrm{FOF}}\sim 60$ which assuming an isothermal density profile, $\rho(r) \propto r^{-2}$, corresponds to an enclosed overdensity with respect to the mean matter density of $\Delta_m \sim 180$. This is close to the value of the virial overdensity predicted by the spherical collapse model in the Einstein-De Sitter cosmology. That is why the percolation parameter is commonly set to $b=0.2$. However, \cite{More2011} have shown that the boundary of FoF halos is not associated to a unique local surface overdensity but is distributed around a characteristic value (for $b=0.2$ this is $\sim 80$). In addition the profile is not isothermal. As a consequence the enclosed overdensity is much higher than 180 and is found to vary in the range $\sim 250$ to $\sim 600$ at $z=0$. It is important to keep this in mind when comparing the properties of N-body halos to observations. In fact, the mass of clusters is usually estimated in terms of the enclosed spherical overdensity with respect to the critical cosmic density that, depending on redshift and cosmology, may correspond to different percolation parameter values. 

\begin{figure}
\includegraphics[width=0.48\textwidth]{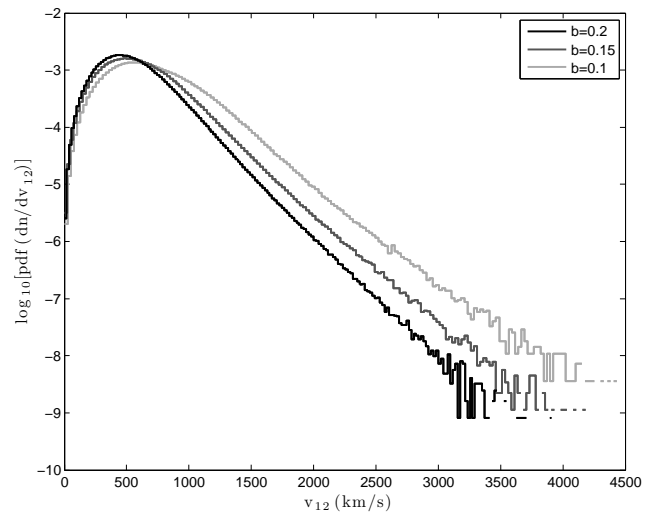}
\caption{Probability density function of the pairwise velocity from the DEUS-FUR $\Lambda$CDM-W7 simulation at $z=0$ for pairs with separation $d_{12} < 15$ h$^{-1}$ Mpc detected assuming linking-length values $b=0.1$ (light grey), $b=0.15$ (grey) and $b=0.2$ (black) respectively.}
\label{bfof_dep}
\end{figure}

In Fig. \ref{bfof_dep} we plot the pairwise velocity distribution $dn/d{\rm v}_{12}$ for halo pairs with a separation $d_{12}<15$ h$^{-1}$Mpc from the DEUS-FUR $\Lambda$CDM-W7 simulation at $z=0$ detected with $b=0.1,0.15$ and $0.2$ respectively (corresponding to a typical variation of the enclosed overdensity $\Delta$ by a factor of $\sim$8). We can see that for increasing values of $b$ the peak of \psc{the distribution increases while its width decreases}. This trend is a direct consequence of the halo definition corresponding to the different values of $b$. Such a difference implies a change of the halo mass function that manifests in the pairwise velocity distribution. 

\begin{figure*}
\includegraphics[width=0.98 \textwidth]{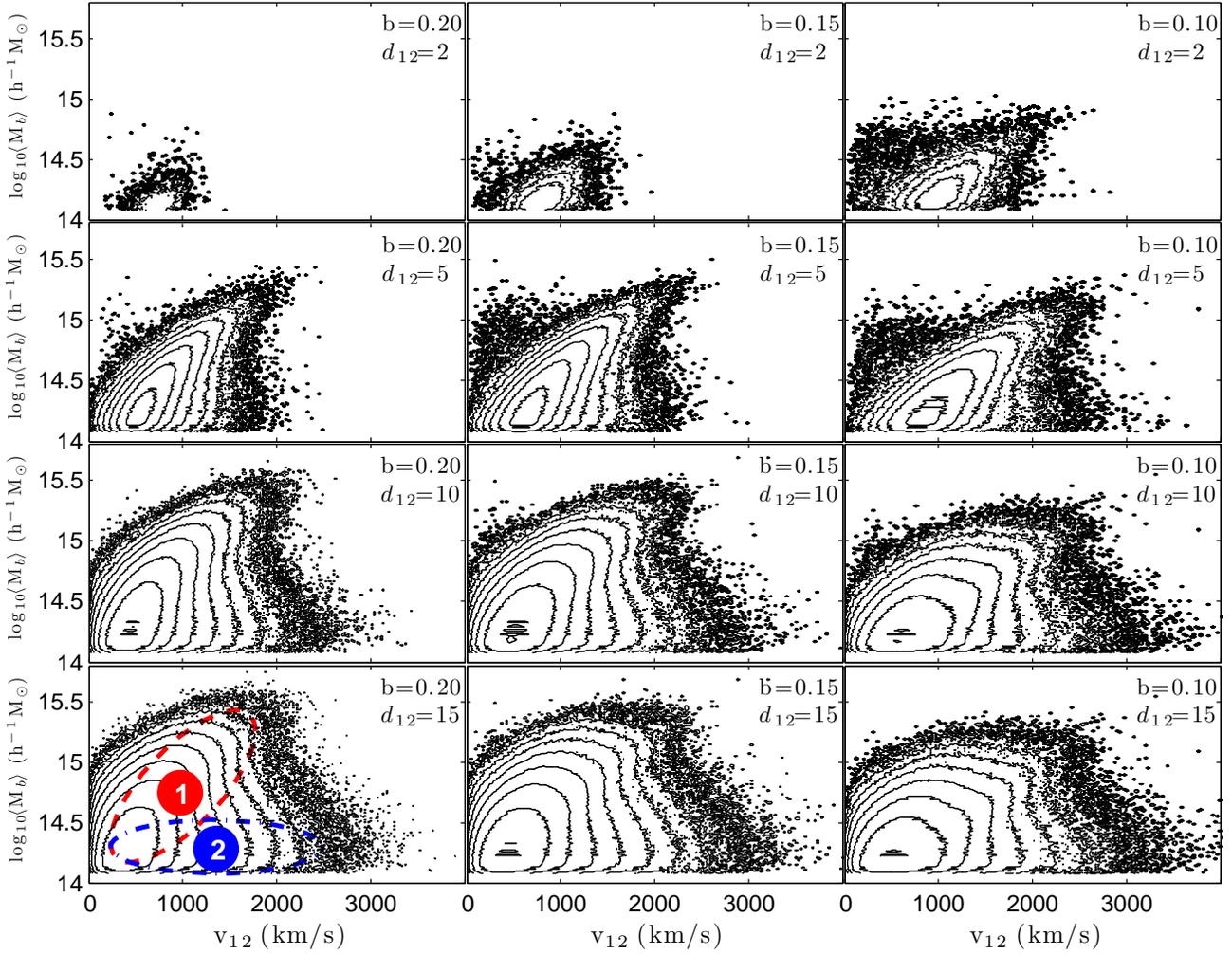}
\caption{Isocontours of the multivariate pairwise velocity probability density function of halo pairs from the DEUS-FUR $\Lambda$CDM-W7 simulation at redshift $z=0$ shown as function of the relative velocity ${\rm v}_{12}$ and the average pair mass $\log_{10} \langle M_b \rangle$, for separation distances $d_{12}=2,5,10$ and $15$ h$^{-1}$Mpc (panels from top to bottom) and linking-length $b=0.2,0.15$ and $0.1$ (panels from left to right) respectively. The isocontours are composed of ten logarithmically spaced bins spanning the range $[10^{-8},10^{-3}]$. For \mbox{$d_{12}<5$ h$^{-1}$Mpc} the pairwise velocity function is characterised by a single population of halos, while for $d_{12}>5$ h$^{-1}$Mpc we notice the emergence of a second population of halo pairs in the low-average-mass range. The former is indicated with Lobe 1 (red dashed ellipse) and its tail consists high-mass low-velocity mergers, the latter which we mark as Lobe 2 (blue dash-dotted line) has a tail consisting of low-mass high-velocity mergers.}
\label{multi_bfof_dep}
\end{figure*}

We may gain a better insight from Fig. \ref{multi_bfof_dep} where we plot isocontours of the multivariate pairwise velocity distribution as function of the relative velocity ${\rm v}_{12}$ and the arithmetic mean mass $\langle $M$_b \rangle$ of halo pairs detected with $b=0.2,0.15$ and $0.1$ (panels from left to right) and distance separation $d_{12}=2,5,10$ and $15$ h$^{-1}$ Mpc (panels from top to bottom) respectively. 

In the case of pairs with $d_{12}=2$ h$^{-1}$ Mpc (top panel) the size of the isocontours increases for decreasing values of $b$. This is because FoF(b=0.1) not only detects individual groups of particles but also sub-structures within halos detected by FoF(b=0.2), thus the former selects a greater number of halo pairs. This population of pairs is distributed in a lobe structure (Lobe 1) which extends from small average masses with low relative velocities to large average masses with moderate relative velocities. The spread of Lobe 1 increases for distance separations $d_{12}>2$ h$^{-1}$ Mpc, simply because FoF detects a greater number of halo pairs. Notice that the tails of the distribution does not exceeds $\approx 2000$ km s$^{-1}$. This indicates that halo pairs in the tail of Lobe 1 correspond to low velocity massive mergers. This could be the case of some observed systems such as MACS J0025.4-1222 \citep{Bradac2008a} or \mbox{ACT-CL J0102-4915} \citep{Menanteau2012a} which \psc{are examples of pairs of massive interacting clusters with Bullet-like baryonic features.}

For $d_{12}>2$ h$^{-1}$ Mpc we can see the emergence of a second lobe (Lobe 2) that causes the multivariate probability density distribution to become bimodal. This second lobe consists of small average mass pairs with velocities extending up to $\approx 3000$ km s$^{-1}$. This is the population that seems to better correspond to the characteristics of 1E0657-56, the Bullet Custer. Notice that the velocity tail of Lobe 2 shifts to larger values for decreasing values of $b$. This is because FoF detects a greater number of \psc{satellite halos that translates into an increase of the number of} small mass high-velocity pairs as $b$ decreases. 

Although the differences of the pairwise velocity probability density function between the case $b=0.15$ and $b=0.2$ might look minor in the high-velocity tail, these may have an important impact on the evaluation of the probability of high-velocity colliding clusters. \psc{Hence, when comparing to observations an important point concern the choice of the percolation parameter which has to be as consistent as possible with the observational mass definition. For instance in \citet{Mastropietro2008a} the colliding halos are spherical objects with a Navarro-Franck-White (NFW) profile \citep{Navarro96,Navarro97} characterized by a concentration parameter $c_{\textrm NFW}\sim 7$ for which the mass is defined in terms of the virial mass $M_{vir}=4\pi/3 r_{vir}^3\Delta_{c}\rho_c$, which is the mass contained in a spherical region of radius $r_{vir}$ enclosing an overdensity $\Delta_{c}=200$ with respect to the critical density of the universe $\rho_c$ at the redshift of the halo. These halos are sampled with $\sim 1000$~particles, thus following \citet{More2011} an enclosed overdensity of $200\rho_c$ at $z\sim 0.5$ roughly corresponds to applying FoF with a linking-length $b\sim 0.15$.}  

\psc{In the following, we therefore adopt a linking-length $b=0.15$ and limit our analysis to halo pairs with distance separation $d_{12}<10$ h$^{-1}$ Mpc which is the minimum distance for which two massive merging halos with virial radius $\sim 5$ h$^{-1}$ Mpc can be detected as distinct objects.}

\section{Statistics of pairwise velocities}\label{sec_cosmoz}
\subsection{Redshift evolution and cosmology dependence}
\psc{We now focus on evaluating the dependence of the pairwise velocity function on redshift and cosmology.} 

\begin{figure}
\includegraphics[width=0.48 \textwidth]{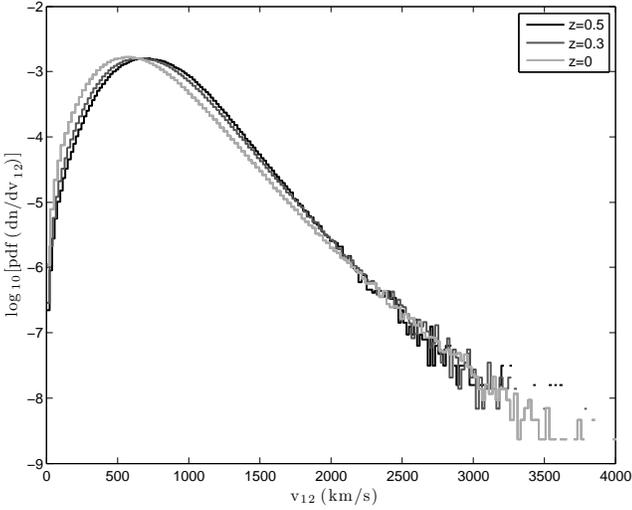}
\caption{Redshift evolution of the probability density function of the pairwise velocity for FoF(b=0.15) halo pairs from the DEUS-FUR $\Lambda$CDM-W7 simulation with distance separation $d_{12} < 10$ h$^{-1}$ Mpc at $z=0.5$ (black), $z=0.3$ (grey) and $z=0$ (light grey) respectively.}
\label{pairwise_redshift}
\end{figure}

\psc{In Fig. \ref{pairwise_redshift} we plot the probability density function associated with $dn/d{\rm v}_{12}$ for halo pairs from the DEUS-FUR $\Lambda$CDM-W7 simulation with distance separation $d_{12}<10$ h$^{-1}$ Mpc at $z=0.5,0.3$ and $0$ respectively. First, we may notice that the amplitude remains constant with redshift, this is the normalization of each curve is different as the number of halo pairs grows with cosmic time. Another effect concerns the tail of the velocity function which tends to slightly increase towards higher velocities from $z=0.5$ to $0$. For instance we find the maximal relative velocities to slowly evolve with redshift with ${\rm v}_{12}^{\rm max}=3609,3799$ and $4000$ km s$^{-1}$ at $z=0.5,0.3$ and $0$ respectively.} Because of this, we can expect that in a given cosmological model the probability (defined as the ratio of the velocity function to the total number of pairs) of finding a halo pair with a large relative velocity at redshift $z=0$ and $z=0.5$ is not significantly different.

\begin{figure}
\includegraphics[width=0.48 \textwidth]{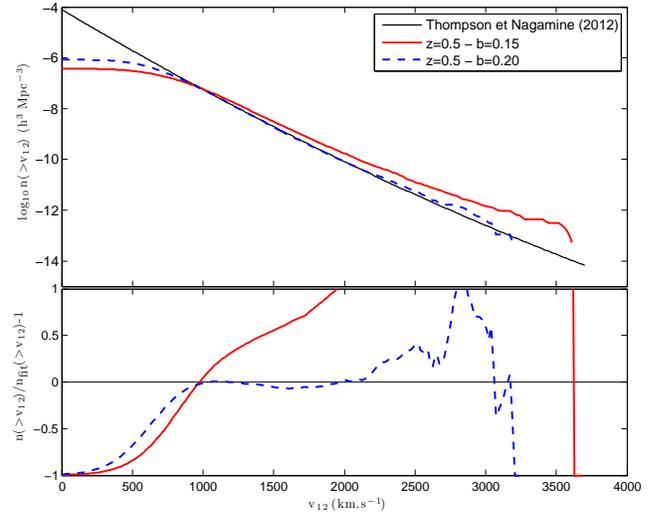}
\caption{Top panel: cumulative pairwise velocity function of halo pairs with distance separation $d_{12} < 10$ h$^{-1}$ Mpc from the DEUS-FUR $\Lambda$CDM-W7 simulation at $z=0.5$ detected with FoF(b=0.2) (blue line) and FoF(b=0.15) (red line). The black solid line corresponds to the quadratic fit from \citet{Thompson2012a}. \psc{Bottom panel: ratio of the cumulative pairwise velocity functions to the \citet{Thompson2012a} fit.}
}
\label{cumulativev12}
\end{figure}

The advantage of using the halo catalogues from the DEUS-FUR simulations can be appreciated from Fig. \ref{pairwise_redshift}. Despite statistical scatter, even at $z=0.5$, the high-velocity tail of the pairwise velocity function is resolved to $v_{12}\approx 3500$ km s$^{-1}$. In this range previous analyses had to strongly rely on extrapolation from fitting functions calibrated to lower relative velocities. For instance, \citet{Thompson2012a} approximated the cumulative pairwise velocity distribution of FoF(b=0.2) halo pairs in $\Lambda$CDM model simulations at $z=0.489$ with a quadratic fit which we plot in the top panel of Fig. \ref{cumulativev12} against the DEUS-FUR $\Lambda$CDM-W7 results. Quite remarkably we can see that it provides a very good approximation up to intermediate velocities.\footnote{The quadratic fit from \citet{Thompson2012a} has been derived from the analysis of FoF(b=0.2) halo pairs in simulations with mass and spatial resolution different from those of the DEUS-FUR simulations. Such differences may affect the halo mass function and introduce a systematic bias in the cumulative pairwise velocity distribution. However, for halo pairs with average masses $>10^{14}\,\textrm{M}_\odot$ this effect remains negligible.} \psc{In contrast, large discrepancies occurs in the high-velocity tail as can be seen from the bottom panel of \mbox{Fig. \ref{cumulativev12}}}. In the same figure we also plot the cumulative distribution from our reference catalogue of halo pairs detected with FoF(b=0.15). As expected from the discussion in Section \ref{dndv_fof} this is characterised by both a greater number of halo pairs and a longer tail at high-velocity compared to $b=0.2$.

\begin{figure}
\includegraphics[width=0.48 \textwidth]{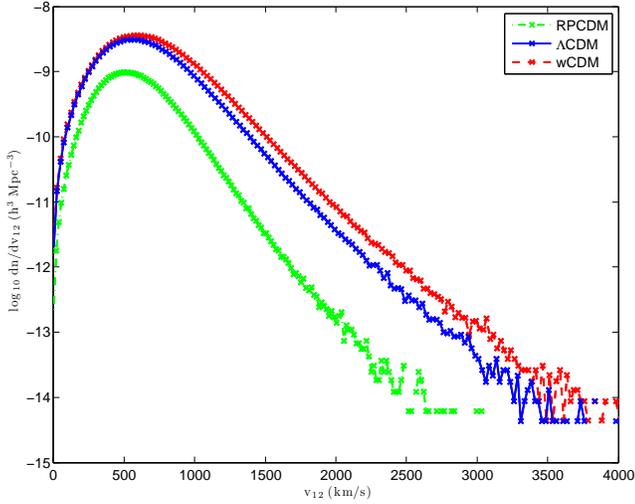}
\caption{Pairwise velocity function at $z=0$ for FoF(b=0.15) halo pairs with distance separation $d_{12} < 10$ h$^{-1}$ Mpc from the DEUS-FUR RPCDM, $\Lambda$CDM-W7 and $w$CDM simulations respectively.
}
\label{cosmoend}
\end{figure}

Let us now turn to the cosmological dependence of the velocity function. In Fig. \ref{cosmoend} we plot $dn/d{\rm v}_{12}$ at $z=0$ for a distance separation $d_{12} < 10$ h$^{-1}$ Mpc for RPCDM, $\Lambda$CDM-W7 and $w$CDM. The first noticeable difference is the overall amplitude of the various curves which is essentially caused by the difference of the mass function of the DEUS-FUR simulated cosmologies. We can also notice that in the high-velocity interval the velocity functions have slightly different slopes, with the $w$CDM and $\Lambda$CDM-W7 models showing a heavier tail than RPCDM. This is indicative of the fact that halo pairs with extreme relative velocities are a sensitive probe of the underlying cosmological model. Understanding the mechanisms responsible for this dependence requires a physical analysis that is beyond the scope of this paper.  

Even in the lack of such study we can have an idea of the dominant cosmological parameter dependence by evaluating the average pairwise velocity $\bar{\rm v}_{12}$. This can be directly inferred from the pairwise probability distribution function obtained from $dn/d{\rm v}_{12}$ and confronted with predictions from the following relation derived in the context of stable clustering \citep{Juszkiewicz1999,Caldwell2001}:
\begin{equation}
-\frac{\bar{\rm v}_{12}(x,a)}{Hr}=\frac{a}{3[1+\xi(x,a)]}\frac{\partial \bar \xi(x,a)}{\partial a} \ ,
\label{eqn1}
\end{equation}
where $\xi$ is the two-point correlation function of the density field, $a$ is the expansion factor, $r=ax$ is the proper separation, $H$ is the Hubble rate and
\begin{equation}
\bar \xi(x,a)=\frac{3}{x^3}\int_0^x \xi(y,a) y^2 dy 
\label{eqn2}
\end{equation}
is the two-point correlation function averaged over a sphere of radius $x$. Evaluating Eqs. (\ref{eqn1}) and (\ref{eqn2}) using the correlation function $\xi$ of the density field of each of the DEUS-FUR simulations, we obtain the following average pairwise velocities: \mbox{$\bar{\rm v}_{12} = 439$ km s$^{-1}$} for the RPCDM model, $\bar{\rm v}_{12} = 490$ km s$^{-1}$ for the $\Lambda$CDM-W7 and $\bar{\rm v}_{12} = 507$ km s$^{-1}$ for the $w$CDM model. These values are within $5\%$ of those directly estimated from the numerical data. Their variation is essentially due to the different values of $\sigma_8$ of the DEUS-FUR cosmologies. This can be understood by taking the ratio of the average pairwise velocity given by Eq. (\ref{eqn1}) for a given $\sigma_8$ with respect to a reference one $\bar{\rm v}_{12}(\sigma_{8,{\rm ref}})$:
\begin{equation}
\frac{\bar{\rm v}_{12} \ (\sigma_8)}{\bar{\rm v}_{12} \ (\sigma_{8,{\rm ref}})}= \frac{\sigma_8}{\sigma_{8,{\rm ref}}} \frac{1+ \xi(r)}{1+\frac{\sigma_8}{\sigma_{8,{\rm ref}}}\xi(r)} \ .\label{eqn3}
\end{equation}
\psc{Assuming that the shape of the power spectrum does not change over the range of scales where the stable clustering regime occurs \citep{Smith2003a} and considering as a reference case the $\Lambda$CDM-W7 values of $\bar{\rm v}_{12}$ and $\sigma_8$, we recover from Eq. \ref{eqn3} the estimated values of the average pairwise velocity of both RPCDM and $w$CDM to better than a few per cent.} 

\begin{figure}
\includegraphics[width=0.48 \textwidth]{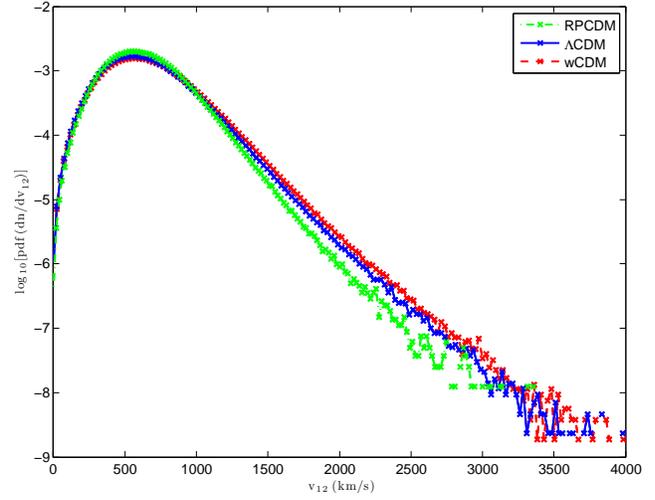}
\caption{Probability density pairwise velocity function at $z=0$ for FoF(b=0.15) halo pairs with distance separation $d_{12} < 10$ h$^{-1}$ Mpc from the DEUS-FUR RPCDM, $\Lambda$CDM-W7 and $w$CDM simulations respectively.}
\label{sigma8_dep}
\end{figure}

\psc{In Fig. \ref{sigma8_dep} we plot the normalized probability density distribution of the pairwise velocity of the three simulated cosmologies. The $\sigma_8$ dependence described above can be seen here on the fact that the distributions have very similar average and overall amplitude. This is because their respective normalizations encode differences of the mass function of the underlying cosmological models that are mostly due to the different $\sigma_8$ values. On the other hand we can see that the high-velocity tail of the distributions is where the cosmological models differ the most. This indicates that the high-velocity tail carry information not only on $\sigma_8$, but also on the cosmic matter density and the properties of the Dark Energy which characterize the simulated models.}

\subsection{Bullet-like halo pairs in DEUS-FUR cosmologies}
In order to identify extreme halo pairs in the DEUS-FUR simulation it is instructive to consider the redshift and cosmology dependence of the multivariate pairwise velocity distribution (see Appendix \ref{Appendix2}). This is shown in Fig. \ref{mvcos} for halo pairs with distance separation $d_{12}<10$ h$^{-1}$ Mpc at $z=0.5,0.3$ and $0$ (panels top to bottom) for DEUS-FUR RPCDM, $\Lambda$CDM-W7 and $w$CDM simulations (panels left to right) respectively. As expected from Section \ref{sec_cosmoz}, for a given cosmological model the range of relative velocities does not vary significantly as function of the redshift compared to interval variation of the average mass of the pairs. Indeed, this is due to the different redshift evolution of the halo mass function compared to that of the pairwise velocity function. Furthermore, we can see that in the RPCDM case the tail of velocity distribution at all redshift remains confined to low velocities. This is not the case of $\Lambda$CDM-W7 and $w$CDM for which we have a large number of pairs with average mass larger than $10^{15.2}$ M$_\odot$ and relative velocity v$_{12}>2100$ km s$^{-1}$. For these models the bimodality of the multivariate pairwise distribution is more pronounced than in RPCDM. In particular, the halo pairs in the tail of Lobe 1 and Lobe 2 of the multivariate distribution clearly indicate that observations of high-mass moderate-velocity merging clusters and low-mass high-velocity ones can provide powerful cosmological probes.

We are especially interested in extremal halos belonging to the latter category. In Appendix \ref{Appendix1} we summarize the characteristics of the highest velocity pairs and the most massive ones detected in the DEUS-FUR simulations. We find that the properties of such extremal pairs vary \psc{with the cosmological model}. As an example, from the values quoted in Table \ref{cosvel} we notice that even at $z=0$ the RPCDM simulation has no pairs with relative velocities exceeding \mbox{$\approx 3000$ km s$^{-1}$} and an average mass $M>2 \times 10^{14}$~h$^{-1}$M$_{\odot}$. \psc{More generally from Tables \ref{cosvel} and \ref{cosmass} we can infer that the deficiency of high-velocity halo pairs with mass above $2 \times 10^{14}$ h$^{-1}$ M$_\odot$ and that of massive pairs with relative velocities above \mbox{$1500$ km s$^{-1}$} tend to disfavor such a cosmological model.} The extremal halo pairs in RPCDM are low-velocity massive mergers with large distance separation. In contrast, in $w$CDM the highest velocity pairs all exceed \mbox{$\approx 4000$ km s$^{-1}$}, though their average masses remain small. 

For comparison, in the top panels of Fig. \ref{mvcos} we plot the average mass and relative velocity of the Bullet Cluster (blue solid lines) as inferred from the analysis of \cite{Mastropietro2008a}. Using this as a reference of the extreme halo pairs we can see that RPCDM has no candidate pairs reproducing the Bullet Cluster characteristics. In the $\Lambda$CDM-W7 case the best candidate pair has a main halo with mass \mbox{$M_1=5.95 \times 10^{14}$ h$^{-1}$ M$_\odot$} and a smaller halo with mass $M_2=1.22 \times 10^{14}$ h$^{-1}$ M$_\odot$ (corresponding to a mass ratio $\sim 5:1$) separated by a distance of 8.4 h$^{-1}$ Mpc, which experience an head on interaction with a relative velocity of 3011 km s$^{-1}$. Notice that such an object is absent from the list of extreme halo pairs shown in Appendix since it has neither an extreme velocity nor a very high mass. In the $w$CDM model the best candidate is characterised by a main halo with a mass $M_1=8.40 \times 10^{14}$ h$^{-1}$ M$_\odot$ and a lighter halo of mass $1.97 \times 10^{14}$ h$^{-1}$ M$_\odot$ (corresponding to a mass ratio of $\sim 4:1$) separated by a distance of 8.8 h$^{-1}$ Mpc and a relative velocity 2839 km s$^{-1}$. \psc{This candidate is no better than that of $\Lambda$CDM. This seems contradictory given the cosmological dependence of the high-velocity tail of the multivariate distribution shown in Fig. \ref{mvcos}. However, such discrepancy can be simply a consequence of the specific realization of the simulation run, such as the phase of the initial conditions. Hence, an object by object comparison is no meaningful in assessing the extremeness of the Bullet Cluster. Such estimation can only be performed through a statistical analysis of extreme halo pairs. In the next Section we will discuss the use of these pairwise velocity catalogs to infer the probability of observing the Bullet Cluster.}

\begin{figure*}
\includegraphics[width=0.98\textwidth]{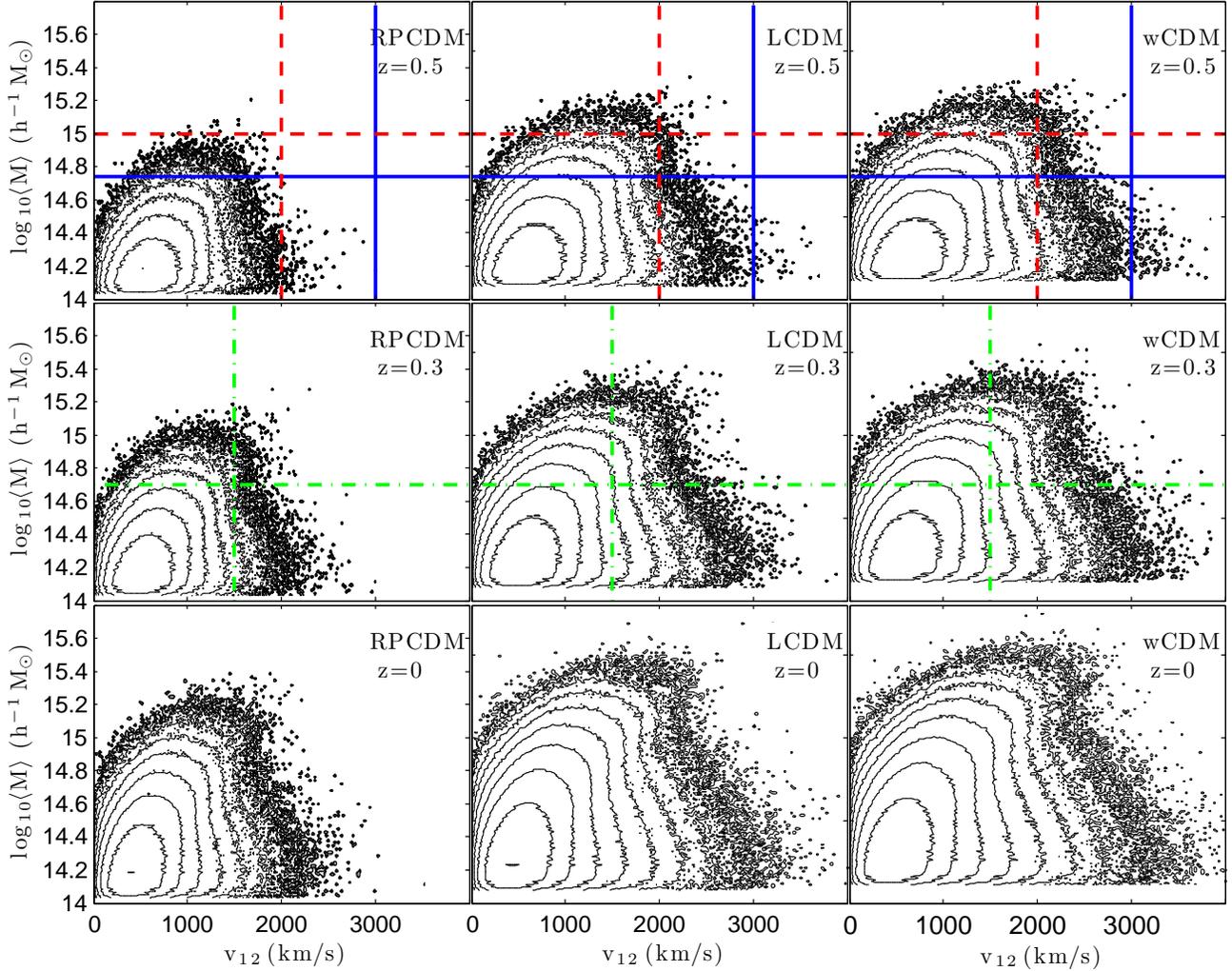}
\caption{Isocontours of the multivariate pairwise velocity probability density function for halo pairs with distance separation $d_{12}<10$ h$^{-1}$Mpc from the DEUS-FUR simulations as function of the relative velocity ${\rm v}_{12}$ and the average pair mass $\log_{10} \langle M \rangle$ for $z=0.5,0.3$ and $0$ (panels from top to bottom) and RPCDM, $\Lambda$CDM-W7 and $w$CDM (panels left to right) respectively. The isocontours are composed of ten logarithmically spaced bins spanning the range $[10^{-8},10^{-3}]$. In the top panels the solid blue lines specify the average mass and relative velocity of the Bullet Cluster from \citep{Mastropietro2008a}. \psc{For illustrative purposes we also show the characteristics of two other Bullet-like systems: the red dashed lines in the top panels corresponds to the characteristics of MACS J0025.4-1222 at the observed redshift of the system \citep{Bradac2008a}, while in the bottom panel the green dot-dashed lines corresponds to the initial characteristics of the Abell 3376 system \citep{Machado2013a}}.
}
\label{mvcos}
\end{figure*}

\section{Extreme Value Statistics of Pairwise Velocities}\label{evsbull}
\subsection{Methodology}
Extreme Value Statistics, originally pioneered by \citet{Frechet1927}, \citet{FisherTippett1928}, \citet{Gumbel1935} and \citet{Gnedenko1943}, has been applied to a wide variety of problems to model the probability of extreme events. Here, we briefly review the basic formalism.

Consider a set of independent identically distributed $N$ random variates $\{X_1,...,X_N\}$ drawn from a cumulative distribution $F(x)$ and $X_\textrm{max}=\textrm{max}\{X_1,...,X_N\}$. It is easy to show that in such case the cumulative distribution function of the maximum of the first $N$ observations is given by
\begin{equation}
P(X_\textrm{max}\le x)=F^N(x). 
\end{equation}
This is the so called \enquote{exact} EVS approach in which $F(x)$ is well known. In the large $N\rightarrow\infty$ limit, it is possible to show that the cumulative distribution function of extreme observations tends to the Generalized Extreme Value (GEV) distribution:
\begin{equation}
P_{[\mu,\sigma,\xi]}(x)=\exp\left\{-\left[1+\xi \left(\frac{x-\mu}{\sigma}\right)\right]^{-1/\xi}\right\},\label{gev}
\end{equation}
defined for $1+\xi(x-\mu)/\sigma>0$, where $\mu$ is the location parameter, $\sigma$ is the scale parameter and $\xi$ is the tail index (or shape parameter), which generalizes the central-limit theorem to extremal subset of data. Depending on the value of $\xi$, Eq. (\ref{gev}) reduces to three possible functional forms: the Gumbel (or type I) distribution ($\xi=0$), the Fr\'echet (or type II) distribution ($\xi>0$) and the Reversed Weibull (or type III) distribution ($\xi<0$). 

Contrary to the exact EVS approach, the use of the Generalized Extreme Value distribution does not require prior knowledge of the underlying cumulative function of the random variates. Instead, it uses these observations to infer the GEV distribution parameters. This is done by classifying the data into blocks of arbitrary size, determining the maxima in each block and inferring the GEV parameters by best fitting the GEV function to the distribution of maxima. A potential disadvantage of this block maxima method is the fact that data need to be sampled. This may cause some loss of valuable rare information or inclusion of non-extremal events. 

A complementary approach that is more suited to our purposes consists in using the Generalized Pareto distribution (GPD). This corresponds to Taylor expanding the tail of Eq. (\ref{gev}) to obtain the cumulative distribution function of observing extreme events above a fixed threshold $\mu$. This reads as 
\begin{equation}
P_{[\mu,\sigma,\xi]}(x)=1- \left [  1+ \xi \left ( \frac{x-\mu}{\sigma} \right)\right ]^{-1/\xi},\label{gpd}
\end{equation}
defined for $1+\xi(x-\mu)/\sigma>0$. Again depending on the value of $\xi$ we have different probabilities of the extreme events. In particular $\xi>0$ ($\xi<0$) corresponds to a long (short) tail distribution. Instead the case $\xi=0$ corresponds to the distribution of events in the tail of a Gaussian distribution. Hence, studies that have extrapolated the probability of the relative velocity of Bullet Cluster in terms of a Gaussian pairwise velocity probability distribution can be seen as a limiting case of the EVS approach described here, with $\xi$ fixed to zero. Notice also that since the tail of the pairwise velocity function depends on the underlying cosmology, we can expect the GPD parameters to carry a strong cosmological dependence. 

\begin{figure}
\includegraphics[width=0.48\textwidth]{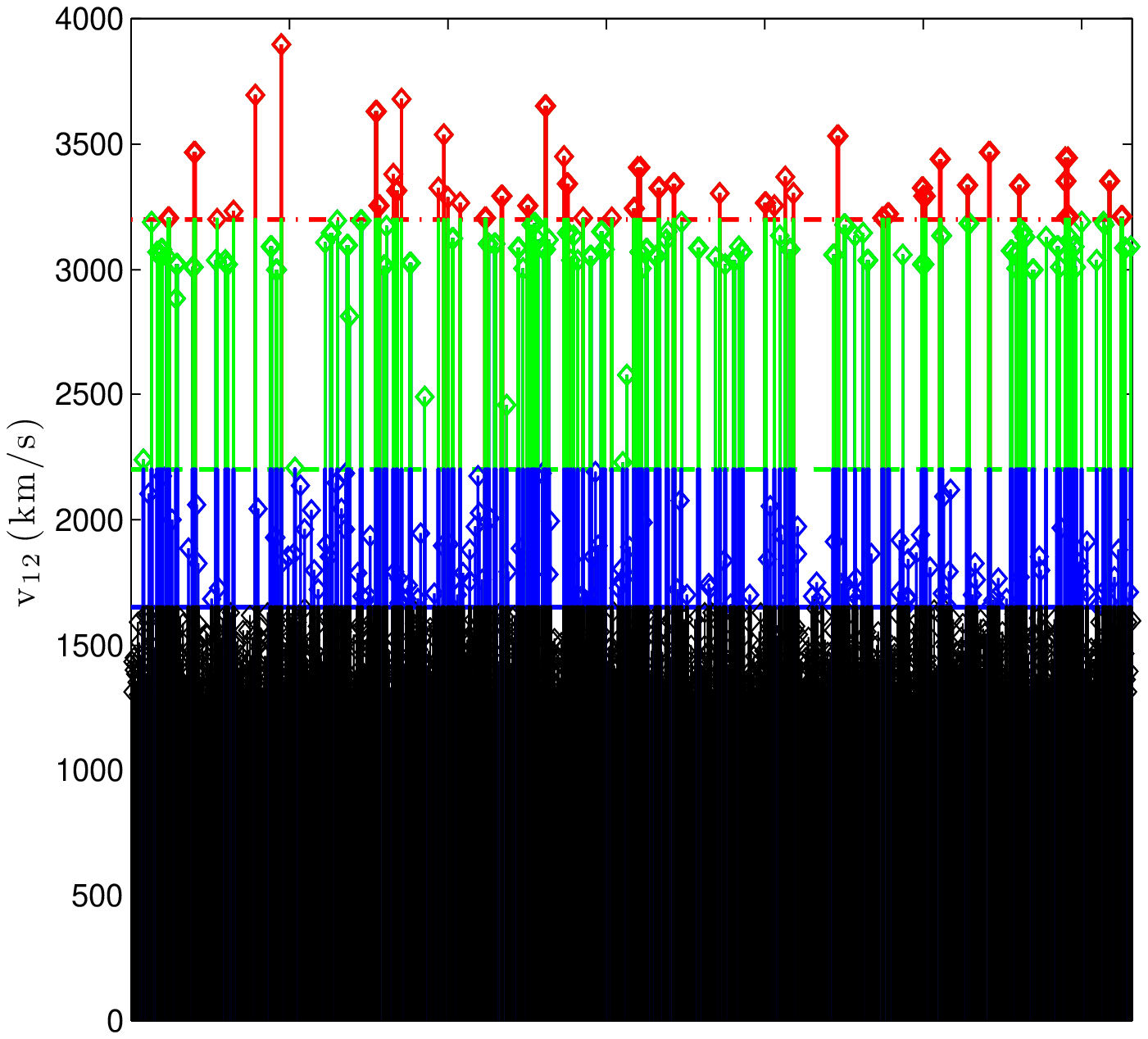}
\caption{Pairwise velocity exceedances defined as the mean of the data minus the chosen threshold for three different thresholds. The corresponding exceedances are shown with different color codes.}
\label{MEF}
\end{figure}

In the GPD approach the issue of sampling the data is replaced by the problem of choosing a suitable value of the threshold. A high threshold would result in a drastic reduction of the data sample, whereas a low threshold may include non-extremal data and thus bias the results toward a gaussian behaviour. This can be seen in Fig. \ref{MEF} for a subset of the pairwise velocities in the DEUS-FUR $\Lambda$CDM-W7 catalog at $z=0.5$ where we have classified the halo pairs according to three different velocity thresholds. Using pairs above the lowest threshold would lead to a gaussian biased estimation of the GPD parameters, while using points above the highest threshold provides a too small sample of extremal events to determine the GPD.

Several statistical diagnostics have been considered to estimate a suitable threshold that segregates common events from extreme ones \citep[for a review see e.g.][]{Scarrott2012}. \psc{Here, we focus on the mean residual life method and the threshold stability plot that have been developed in relation with EVS data analysis problems.} 

The mean residual life method consists in plotting the mean excess, defined as the mean of the exceedances of the data minus the threshold, as function of the threshold itself. An optimal choice is then given by the lowest threshold value for which all higher thresholds give a sequence of mean excesses that is consistent with a straight line \citep[][]{Scarrott2012}. 

\begin{figure}
\includegraphics[width=0.48\textwidth]{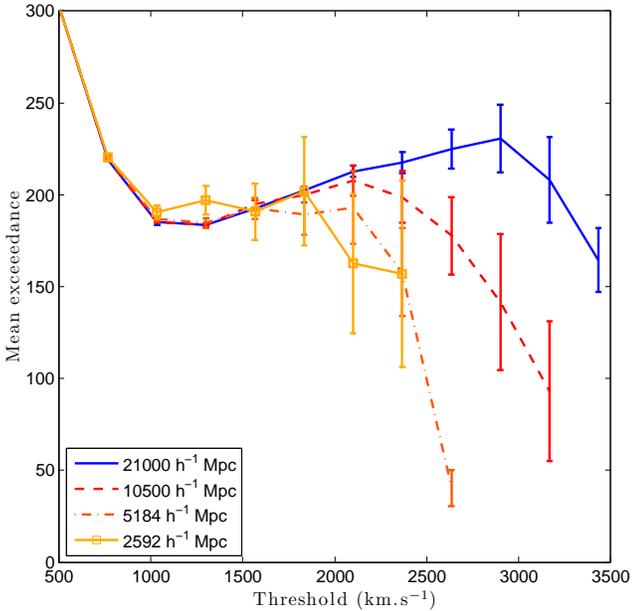}
\caption{\psc{Mean residual life plot of the mean excess as function of the threshold for the full DEUS-FUR $\Lambda$CDM-W7 simulation volume (blue solid line) and three subvolumes of boxlength $10500$ h$^{-1}$ Mpc (red dashed line), $5184$ h$^{-1}$ Mpc (orange dot-dashed line) and \mbox{$2592$ h$^{-1}$ Mpc} (yellow squared solid line) respectively. For the full volume the region of stability, where the mean excess evolves as a straight line, spans the range \mbox{$\sim 1500$ km s$^{-1}$} to \mbox{$\sim 2500$ km s$^{-1}$}. The selected threshold \mbox{$\mu=2100$ km s$^{-1}$} corresponds to the largest threshold value with smallest statistical errors. We can see that for decreasing volumes the stability region rapidly shrinks while becoming more uncertain such that for small simulation volumes it is not possible to reliably select a threshold for which the results inferred from the GPD remain stable}.
}
\label{MED}
\end{figure}

\psc{A mean residual life plot is shown in Fig. \ref{MED} for the halo pairs of the full DEUS-FUR $\Lambda$CDM-W7 simulation volume (blue solid line) and three subvolumes of boxlength $10500$ h$^{-1}$ Mpc (red dashed line), $5184$ h$^{-1}$ Mpc (orange dot-dashed line) and \mbox{$2592$ h$^{-1}$ Mpc} (yellow squared solid line) respectively. We can see a characteristic trend with the mean of the exceedances rapidly decreasing at low threshold values, while increasing linearly at intermediate thresholds and then sharply decreasing at large values. The main difference among the various volume catalogs is the interval extent and statistical uncertainty of the linear trend. In the case of the full DEUS-FUR volume this interval has the maximum extent implying a precise selection of the GPD threshold which also guarantee a stability of GPD inferred results. For smaller volumes the interval ranges is much smaller and more uncertain such that it becomes impossible to reliably select a threshold value.}

\begin{figure}
\includegraphics[width=0.48\textwidth]{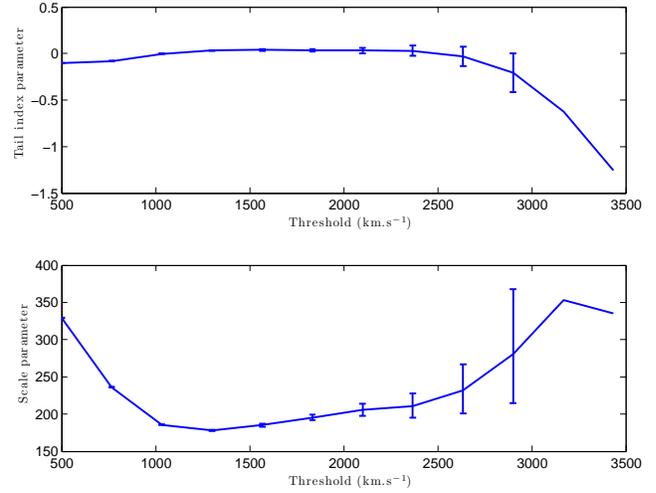}
\caption{Tail index (top panel) and scale parameter (bottom panel) diagnostic plots for the generalized Pareto distribution distribution fit to the $\Lambda$CDM-W7 data as function of the threshold parameter. 
}
\label{fig:stability}
\end{figure}

The threshold stability plot is joint diagnostic that consists in plotting the GPD shape and scale parameter values best fitting the data as function of the threshold. Then, an optimal threshold value is chosen such that for higher values the GPD parameters remain stable \citep[][]{Scarrott2012}. We show such a plot in Fig. \ref{fig:stability}. 

From Fig. \ref{MED} and Fig. \ref{fig:stability} we can see that the curves are nearly constant straight lines in the threshold range $2000$ to \mbox{$2500$ km s$^{-1}$}, thus we set $\mu=2100$ km s$^{-1}$. This guarantees the stability of the results with respect to the choice of the threshold. Performing a similar analysis for the other DEUS-FUR cosmological simulations we set $\mu=1870$ km s$^{-1}$ for the RPCDM case and \mbox{$\mu=2151$ km s$^{-1}$} for the $w$CDM model respectively. 

In Fig. \ref{GPfitting} we plot the Generalized Pareto density distributions best fitting the probability density distribution functions of halo pairs with distance separation $d_{12} < 10$ h$^{-1}$ Mpc and average pair mass $>$10$^{14}$ h$^{-1}$ M$_\odot$ from the DEUS-FUR cosmological simulation catalogs at $z=0.5$. \psc{The selected thresholds for the different cosmological models are indicated by vertical dashed lines, while the dot-dashed lines correspond to the Gaussian tails (with $\xi=0$) for the same threshold and scale parameter values.}

\psc{The best-fit values and the $68\%$ confidence interval of $\xi$ and $\sigma$ have been determined through a Monte Carlo Markov Chain likelihood analyses of the binned numerical data assuming Poisson errors. The inferred values are quoted in Table \ref{ParetoPar}. Notice that in all cases a Gaussian tail ($\xi=0$) is excluded at more than $99.7\%$ confidence level. Since the best-fit value of the shape parameter is positive this implies the probability distribution of extreme pairwise velocities is slightly heavy tailed, which increases the probability of finding high relative velocity pairs compared to previous studies that have simply assumed a Gaussian distribution \citep{Lee2010a,Thompson2012a}.}

\subsection{Application to the Bullet Cluster}
\begin{figure}
\includegraphics[width=0.48\textwidth]{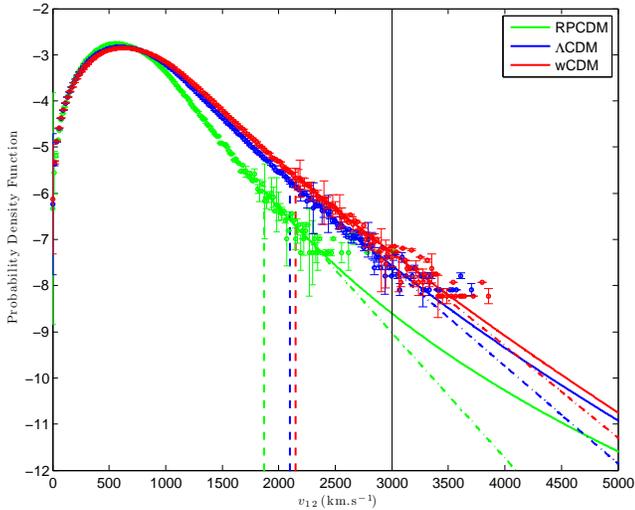}
\caption{Probability density distribution of the pairwise velocity of halo pairs with average mass $>$10$^{14}$ h$^{-1}$ M$_\odot$ and distance separation \mbox{$d_{12} < 10$ h$^{-1}$ Mpc} from the DEUS-FUR simulations at $z=0.5$ for $\Lambda$CDM-W7 (blue points), $w$CDM scenario (red points) and RPCDM (green points) models respectively. The solid lines shows the tail of the Generalized Pareto distributions best fitting the numerical data with threshold values indicated by the vertical dashed lines. \psc{The dot-dashed lines corresponds to the Gaussian tails with threshold and scale parameters set to that of the best-fit GPD tails of the three cosmologies.} The thin solid black line at \mbox{v$_{12}=3000$ km s$^{-1}$} corresponds to the Bullet Cluster relative velocity estimated in \citet{Mastropietro2008a}.
}
\label{GPfitting}
\end{figure}

Having determined the GPD parameters from each of the DEUS-FUR halo pair catalogs, we are now able to estimate the probability of observing the Bullet Cluster for different DEUS-FUR cosmological models. This is obtained by integrating the probability density functions shown in Fig. \ref{GPfitting} from $\textrm{v}_{12}=3000\,{\rm km \,s^{-1}}$ to infinity. This probability has to be interpreted as the rate of occurrence of Bullet Cluster-like systems in comoving space.

In the $\Lambda$CDM-W7 case we find $P({\rm v}_{12}>3000\,{\rm km\,s^{-1}})=6.4\times 10^{-6}$, which is two orders of magnitude larger than previous estimates \citep{Lee2010a,Thompson2012a}. For the RPCDM model we obtain $P({\rm v}_{12}>3000\,{\rm km \,s^{-1}})=9.7\times 10^{-8}$, while we find $P({\rm v}_{12}>3000\,{\rm km \,s^{-1}})=1.7\times 10^{-5}$ for the $w$CDM case. 
 
As shown in Section \ref{sec_cosmoz} the pairwise velocity distribution carries information on cosmological model parameters such as $\sigma_8$, as well as $\Omega_m$ and $w$ which differentiate the DEUS-FUR cosmologies. The value of these parameters have been selected along the $\sigma_8-w$ (and $\Omega_m-w$) degeneracy line of the CMB (and SN Ia) data. Henceforth, the Bullet Cluster inferred probabilities can be used to provide us with some qualitative constraints on these class of models. In particular, these suggest that the observation of the Bullet Cluster strongly disfavors Dark Energy models, such as RPCDM, which have an equation of state $w>-1$ for which CMB data enforce smaller $\sigma_8$ values to compensate for the greater amplitude of the integrated Sachs-Wolfe \citep{ISW} effect on the CMB temperature anisotropy power spectrum \citep{Kunzetal2004} while SN Ia data enforce a lower value of $\Omega_m$ to compensate for the shorter luminosity distance. These models are characterized by a lower level of matter clustering with respect to the standard $\Lambda$CDM-W7 model. In contrast, the probability of finding the Bullet Cluster increases in the case of Dark Energy models with more negative values of the equation of state $w\le -1$ for which CMB data enforces larger $\sigma_8$ values while the SN Ia data requires larger values of $\Omega_m$. These models are characterized by a higher level of matter clustering compared to the $\Lambda$CDM-W7 case. Henceforth, it is plausible that the statistical measurements of the rate of occurrence of bullet cluster-like systems which sample the tail of the pairwise velocity distribution have the potential probe Dark Energy and break degeneracy lines of the underlying cosmological parameters.

\begin{table}
\begin{center}
\begin{tabular}{cccc}
\hline
Parameters & RPCDM & $\Lambda$CDM-W7 & $w$CDM \\
\hline
\hline
$\xi$ & $0.073 \pm 0.010$ & $0.035 \pm 0.007$ & $0.020 \pm 0.008$ \\
$\sigma$ (km s$^{-1}$) & $159.4 \pm 0.5$ & $205.1 \pm 0.4$ & $218.3 \pm 0.4$ \\
\hline
$\mu$ (km s$^{-1}$) & $1870$ & $2100$ & $2151$ \\
\hline
\end{tabular}
\caption{Best-fit values and $1\sigma$ errors of the GPD parameters for the different DEUS-FUR cosmological models: $\xi$ is the tail index parameter, $\sigma$ is the scale parameter and $\mu$ is the threshold. \psc{The constraints on $\xi$ indicates deviations from a Gaussian distribution at more than $99.7\%$ confidence level}. 
}
\label{ParetoPar}
\end{center}
\end{table}

\section{Conclusion}\label{conclu}
We have explored the possibility of testing cosmological models through observations of extreme pairwise velocities of interacting galaxy clusters. To this purpose we have studied the properties of pairwise velocities from the halo catalogs of the DEUS-FUR cosmological simulations. Thanks to the large simulation volume we have been able to resolve the high-velocity tail of the pairwise velocity distribution. We have studied its dependence on the percolation parameter of the FoF halo finder, the distance separation, the redshift evolution and cosmology. We have shown that a particular attention has to be paid to the halo mass definition, since the choice of the percolation parameter especially alter the tail of pairwise velocity function. In the redshift range $z=0$ to $0.5$ the latter show minor evolution, while it significantly varies with cosmology. To have an idea of the cosmological model parameter dependence we have estimated the average pairwise velocity of the DEUS-FUR cosmologies using a model based on stable clustering. From the comparison with the mean value inferred from the DEUS-FUR halo pairs catalogs we have shown that most of the average cosmological dependence is driven by the value of $\sigma_8$ while the tail of the distribution carries information on $\sigma_8$, $\Omega_m$ and $w$ which differentiate the DEUS-FUR cosmologies. As such observations of extreme relative velocities can be used \vb{as a different probe to measure the equation of state of Dark Energy and test cosmological models}. In particular, the analysis of the multivariate pairwise velocity distribution indicates that observations of low-mass high-velocity interacting clusters (e.g. Bullet Cluster) as well as massive systems with moderate relative velocities are most sensitive to the underlying cosmology.

Focusing on the Bullet Cluster system, we have found a number of halo pairs candidates in $\Lambda$CDM-W7 and $w$CDM catalogs respectively, while we have found none in the RPCDM case within the simulated volume of the observable universe. Built upon these results we have quantified the probability of observing the Bullet Cluster in the context of Extreme Value Statistics. To this end we have used Generalized Pareto distributions to model the probability distribution of the pairwise velocities. We find the probability of observing a halo pair with average mass $>$10$^{14}$ h$^{-1}$ M$_\odot$, distance separation \mbox{$d_{12} < 10$ h$^{-1}$ Mpc} and relative velocity $>3000$ km s$^{-1}$ to strongly vary across the DEUS-FUR simulated cosmologies with probabilities of $9.7\times 10^{-8}$, $6.4\times 10^{-6}$ $1.7\times 10^{-5}$ for RPCDM, $\Lambda$CDM-W7 and $w$CDM respectively. Thus, we can deduce that the observation of the Bullet Cluster strongly disfavours cosmologies with low value of the $\sigma_8$ (low level of matter clustering). In the case of Dark Energy models calibrated against CMB observations this occurs for $w>-1$ since CMB data enforce low $\sigma_8$ values primarily to compensate for the enhanced amplitude of the ISW effect. In contrast, the probability of the Bullet Cluster increases for models with larger $\sigma_8$ and thus a higher level of matter clustering.

The study presented here suggests that observations of extreme interactive clusters sampling the tail of the pairwise velocity distribution can provide complementary information on Dark Energy models potentially capable of breaking standard cosmological parameter degeneracies. A first step in this direction will be the development of an accurate theoretical model of the pairwise velocity distribution density function along the line of the original work by \citep{Sheth1996,DiaferioSheth2001}. This will help elucidating the cosmological dependence of the high-velocity tail and provide an estimate of the number of bullet-like systems in the sky that need to be observed to improve current parameter constraints on Dark Energy.

\section*{Acknowledgements}
We warmly thank Vincent Reverdy and Ir\`ene Balmes for fruitful discussions about EVS and their deep implication in the achievement of the DEUS-FUR project. This work was granted access to HPC resources of TGCC through allocations made by GENCI (Grand \'Equipement National de Calcul Intensif) in the framework of the \enquote{Grand Challenge} DEUS. The research leading to these results has received funding from the European Research Council under the European Union's Seventh Framework Programme (FP/2007-2013) / ERC Grant Agreement n. 279954. We acknowledge support from the DIM ACAV of the Region Ile de France. V.~R. Bouillot is supported financially (AW) by the National Research Foundation of South Africa. Any opinion, findings and conclusions or recommendations expressed in this material are those of the authors and therefore the NRF does not accept any liability in regard thereto.

\appendix

\section{Properties of Extremal Halo Pairs in DEUS-FUR simulations}
\label{Appendix1}
In this Appendix, we present the characteristics of pairs of halos detected in the DEUS-FUR simulations at different redshifts for different cosmological models. For each pair we quote the relative velocity v$_{12}$, the mass of the main halo $M_1$ and the satellite $M_2$, the mass ratio, the distance separation and the colliding angle. We only consider pairs with distance separation $d_{12}<10$ h$^{-1}$ Mpc. 

\begin{table*}
\begin{minipage}{150mm}
\begin{center}
\caption{Characteristics of the highest velocity pairs in the DEUS-FUR $\Lambda$CDM-W7 model simulation. Haloes are detected with a linking length $b=0.15$ and the maximum separation for the pair is $d_{12}<10$ h$^{-1}$ Mpc.}
\label{velLCDMredshift}
\begin{tabular}{ccccccc}
\hline
 & v$_{12}$ & M$_1 ($h$^{-1}$ M$_\odot$) & M$_2 ($h$^{-1}$ M$_\odot$) & M$_1/$M$_2$ & $d_{12} ($h$^{-1}$ Mpc) & $\theta$  \\
\hline
$z=0$ & $N_{\rm pairs}=71,454,161$ \\
& 4000 & 1.46 $\times 10^{14}$ & 1.24 $\times 10^{14}$ & 1.18 & 9.57 & 19 \\
& 3845 & 2.20 $\times 10^{14}$ & 1.28 $\times 10^{14}$ & 1.71 & 6.70 & 3 \\
& 3835 & 1.88 $\times 10^{14}$ & 1.62 $\times 10^{14}$ & 1.16 & 9.39 & 7 \\
& 3760 & 1.76 $\times 10^{14}$ & 1.46 $\times 10^{14}$ & 1.20 & 9.05 & 25 \\
& 3756 & 5.96 $\times 10^{14}$ & 1.68 $\times 10^{14}$ & 3.55 & 8.98 & 6 \\
$z=0.3$ & $N_{\rm pairs}=27,923,366$ \\
& 3790 & 1.39 $\times 10^{14}$ & 1.30 $\times 10^{14}$ & 1.07 & 6.52 & 9 \\
& 3498 & 1.76 $\times 10^{14}$ & 1.26 $\times 10^{14}$ & 1.40 & 7.91 & 11 \\
& 3306 & 1.74 $\times 10^{14}$ & 1.25 $\times 10^{14}$ & 1.39 & 9.01 & 12 \\
& 3301 & 3.90 $\times 10^{14}$ & 2.96 $\times 10^{14}$ & 1.31 & 9.77 & 25 \\
& 3265 & 2.52 $\times 10^{14}$ & 1.24 $\times 10^{14}$ & 2.03 & 5.17 & 13 \\
   $z=0.5$& $N_{\rm pairs}=13,101,859$ \\
& 3609 & 3.17 $\times 10^{14}$ & 1.25 $\times 10^{14}$ & 2.53 & 8.74 & 3 \\
& 3587 & 2.56 $\times 10^{14}$ & 1.73 $\times 10^{14}$ & 1.47 & 7.03 & 11 \\
& 3543 & 4.09 $\times 10^{14}$ & 1.36 $\times 10^{14}$ & 3.01 & 8.70 & 13 \\
& 3425 & 1.75 $\times 10^{14}$ & 1.61 $\times 10^{14}$ & 1.09 & 5.91 & 13 \\
& 3271 & 3.59 $\times 10^{14}$ & 2.28 $\times 10^{14}$ & 1.57 & 7.93 & 8 \\
\\
\hline
\end{tabular}
\end{center} 
\end{minipage}
\end{table*}

\begin{table*}
\begin{minipage}{150mm}
\begin{center}
\caption{Characteristics of the highest mass pairs in the DEUS-FUR $\Lambda$CDM-W7 model simulation. Haloes are detected with a linking length $b=0.15$ and the maximum separation for the pair is $d_{12}<10$ h$^{-1}$ Mpc.}
\label{massLCDMredshift}
\begin{tabular}{ccccccc}
\hline
 & v$_{12}$ & M$_1 ($h$^{-1}$ M$_\odot$) & M$_2 ($h$^{-1}$ M$_\odot$) & M$_1/$M$_2$ & $d_{12} ($h$^{-1}$ Mpc) & $\theta$  \\
\hline
$z=0$ & $N_{\rm pairs}=71,454,161$ \\
& 3227 & 96.1 $\times 10^{14}$ & 2.45 $\times 10^{14}$ & 39.27 & 5.88 & 35 \\
& 2902 & 96.1 $\times 10^{14}$ & 1.36 $\times 10^{14}$ & 70.89 & 7.66 & 5 \\
& 2169 & 63.7 $\times 10^{14}$ & 6.74 $\times 10^{14}$ & 9.45 & 6.51 & 2 \\
& 2306 & 69.1 $\times 10^{14}$ & 1.27 $\times 10^{14}$ & 54.31 & 5.59 & 9 \\
& 2323 & 64.2 $\times 10^{14}$ & 1.60 $\times 10^{14}$ & 40.25 & 6.35 & 17 \\
$z=0.3$ & $N_{\rm pairs}=27,923,366$ \\
& 1901 & 54.2 $\times 10^{14}$ & 1.81 $\times 10^{14}$ & 29.91 & 4.32 & 17 \\
& 2300 & 46.3 $\times 10^{14}$ & 6.06 $\times 10^{14}$ & 7.63 & 9.67 & 15 \\
& 2202 & 46.3 $\times 10^{14}$ & 1.72 $\times 10^{14}$ & 26.97 & 6.30 & 9 \\
& 2584 & 46.2 $\times 10^{14}$ & 1.64 $\times 10^{14}$ & 28.09 & 5.71 & 12 \\
& 1771 & 43.6 $\times 10^{14}$ & 3.77 $\times 10^{14}$ & 11.56 & 9.06 & 4 \\
   $z=0.5$& $N_{\rm pairs}=13,101,859$ \\
& 2337 & 42.8 $\times 10^{14}$ & 1.88 $\times 10^{14}$ & 22.73 & 4.71 & 3 \\
& 2093 & 35.7 $\times 10^{14}$ & 3.10 $\times 10^{14}$ & 11.52 & 6.93 & 9 \\
& 2035 & 28.0 $\times 10^{14}$ & 7.63 $\times 10^{14}$ & 3.67 & 8.2 & 8 \\
& 2077 & 27.2 $\times 10^{14}$ & 8.47 $\times 10^{14}$ & 3.20 & 7.68 & 4 \\
& 2078 & 31.1 $\times 10^{14}$ & 2.60 $\times 10^{14}$ & 11.96 & 7.18 & 4 \\
\hline
\end{tabular}
\end{center} 
\end{minipage}
\end{table*}

\begin{table*}
\begin{minipage}{150mm}
\begin{center}
\caption{Characteristics of the highest velocity pairs in the three DEUS-FUR cosmologies at $z=0$. Haloes are detected with a linking length $b=0.15$ and the maximum separation for the pair is $d_{12}<10$ h$^{-1}$ Mpc.}
\label{cosvel}
\begin{tabular}{ccccccc}
\hline
 & v$_{12}$ & M$_1 ($h$^{-1}$ M$_\odot$) & M$_2 ($h$^{-1}$ M$_\odot$) & M$_1/$M$_2$ & $d_{12} ($h$^{-1}$ Mpc) & $\theta$  \\
\hline
RPCDM & $N_{\rm pairs}=17,579,037$ \\
& 3037 & 2.75 $\times 10^{14}$ & 1.20 $\times 10^{14}$ & 2.29 & 9.88 & 14 \\
& 3011 & 1.21 $\times 10^{14}$ & 1.21 $\times 10^{14}$ & 1.00 & 6.18 & 4 \\
& 2844 & 1.84 $\times 10^{14}$ & 1.32 $\times 10^{14}$ & 1.39 & 9.08 & 4 \\
& 2833 & 1.99 $\times 10^{14}$ & 1.16 $\times 10^{14}$ & 1.72 & 8.23 & 7 \\
& 2767 & 3.59 $\times 10^{14}$ & 1.62 $\times 10^{14}$ & 2.21 & 9.42 & 10 \\
$\Lambda$CDM-W7  & $N_{\rm pairs}=71,454,161$ \\
& 4000 & 1.46 $\times 10^{14}$ & 1.24 $\times 10^{14}$ & 1.18 & 9.57 & 19 \\
& 3845 & 2.20 $\times 10^{14}$ & 1.28 $\times 10^{14}$ & 1.71 & 6.70 & 3 \\
& 3835 & 1.88 $\times 10^{14}$ & 1.62 $\times 10^{14}$ & 1.16 & 9.39 & 7 \\
& 3760 & 1.76 $\times 10^{14}$ & 1.46 $\times 10^{14}$ & 1.20 & 9.05 & 25 \\
& 3756 & 5.96 $\times 10^{14}$ & 1.68 $\times 10^{14}$ & 3.55 & 8.98 & 6 \\
$w$CDM & $N_{\rm pairs}=90,232,273$ \\
& 4923 & 1.84 $\times 10^{14}$ & 1.64 $\times 10^{14}$ & 1.126 & 8 & 5 \\
& 4357 & 3.32 $\times 10^{14}$ & 1.35 $\times 10^{14}$ & 2.448 & 7.35 & 8 \\
& 4166 & 3.82 $\times 10^{14}$ & 1.34 $\times 10^{14}$ & 2.846 & 9.66 & 4 \\
& 4112 & 2.85 $\times 10^{14}$ & 1.37 $\times 10^{14}$ & 2.085 & 9.1 & 26 \\
& 4031 & 1.48 $\times 10^{14}$ & 1.29 $\times 10^{14}$ & 1.15 & 9.28 & 12 \\
\hline
\end{tabular}
\end{center} 
\end{minipage}
\end{table*}

\begin{table*}
\begin{minipage}{150mm}
\begin{center}
\caption{Characteristics of the highest mass pairs in the three DEUS-FUR cosmologies at $z=0$. Haloes are detected with a linking length $b=0.15$ and the maximum separation for the pair is $d_{12}<10$ h$^{-1}$ Mpc.}
\label{cosmass}
\begin{tabular}{ccccccc}
\hline
 & v$_{12}$ & M$_1 ($h$^{-1}$ M$_\odot$) & M$_2 ($h$^{-1}$ M$_\odot$) & M$_1/$M$_2$ & $d_{12} ($h$^{-1}$ Mpc) & $\theta$  \\
\hline
RPCDM & $N_{\rm pairs}=17,579,037$ \\
& 1890 & 43.8 $\times 10^{14}$ & 5.91 $\times 10^{14}$ & 7.41 & 9.69 & 13 \\
& 1733 & 43.8 $\times 10^{14}$ & 2.25 $\times 10^{14}$ & 19.50 & 5.49 & 4 \\
& 1385 & 42.9 $\times 10^{14}$ & 1.95 $\times 10^{14}$ & 21.94 & 8.43 & 19 \\
& 1278 & 42.9 $\times 10^{14}$ & 1.89 $\times 10^{14}$ & 22.69 & 9.24 & 23 \\
& 1440 & 40.0 $\times 10^{14}$ & 3.18 $\times 10^{14}$ & 12.61 & 9.01 & 4 \\
$\Lambda$CDM-W7 & $N_{\rm pairs}=71,454,161$ \\
& 3227 & 96.1 $\times 10^{14}$ & 2.45 $\times 10^{14}$ & 39.27 & 5.88 & 35 \\
& 2902 & 96.1 $\times 10^{14}$ & 1.36 $\times 10^{14}$ & 70.89 & 7.66 & 5 \\
& 2169 & 63.7 $\times 10^{14}$ & 6.74 $\times 10^{14}$ & 9.45 & 6.51 & 2 \\
& 2306 & 69.1 $\times 10^{14}$ & 1.27 $\times 10^{14}$ & 54.31 & 5.59 & 9 \\
& 2323 & 64.2 $\times 10^{14}$ & 1.60 $\times 10^{14}$ & 40.25 & 6.35 & 17 \\
$w$CDM & $N_{\rm pairs}=90,232,273$ \\
& 2665 & 74.7 $\times 10^{14}$ & 17.3 $\times 10^{14}$ & 4.32 & 7.62 & 13 \\
& 2295 & 65.9 $\times 10^{14}$ & 22.8 $\times 10^{14}$ & 2.88 & 9.70 & 3 \\
& 2251 & 85.1 $\times 10^{14}$ & 1.50 $\times 10^{14}$ & 56.87 & 8.76 & 11 \\
& 2308 & 80.9 $\times 10^{14}$ & 3.03 $\times 10^{14}$ & 26.69 & 9.21 & 9 \\
& 1800 & 76.1 $\times 10^{14}$ & 6.06 $\times 10^{14}$ & 12.56 & 9.95 & 18 \\
\hline
\end{tabular}
\end{center} 
\end{minipage}
\end{table*}

\section{Details of the multivariate distribution}
\label{Appendix2}
In this Appendix, we present the maximum relative velocity or the maximum average mass of the pairs of halos detected in the DEUS-FUR simulations at different redshifts for different cosmological models having set a prior on the minimum mass (Table \ref{table_multi1}) or on the minimum relative velocity (Table \ref{table_multi2}). This table selects halos following the same procedure as in section \ref{evsbull} and aims at highlighting some Bullet Cluster candidates in the three DEUS-FUR cosmologies.
For each sample we quote the maximal relative velocity v$_{12}$. We only consider pairs with distance separation $d_{12}<10$ h$^{-1}$ Mpc. 

\begin{table*}
\begin{minipage}{150mm}
\begin{center}
\caption{Pairwise velocity (in km s$^{-1}$) of the fastest halo pairs with average mass above three threshold values for the DEUS-FUR cosmologies at z = 0, 0.3 and 0.5 respectively. Haloes are detected with a linking length $b=0.15$ and the maximum separation for the pair is $d_{12}<10$ h$^{-1}$ Mpc.}
\label{table_multi1}
\begin{tabular}{clccccc}
\hline
 & $<$M$> \ge$ & M$_\textrm{min}$ (h$^{-1}$ M$_\odot$)  & $3 \times 10^{14}$ h$^{-1}$ M$_\odot$  & $5 \times 10^{14}$ h$^{-1}$ M$_\odot$ & $10^{15}$ h$^{-1}$ M$_\odot$  \\
\hline
RPCDM & $z=0$&   3528  &   2953 &   2657 &   2230 \\
 & $z=0.3$&   3008  &   2697 &   2309 &   2309  \\
 & $z=0.5$&   3010  &   2256 &   2042 &   1803  \\
\hline
$\Lambda$CDM-W7 &$z=0$ &   4954  &   3929 &   3444 &   3384\\
 & $z=0.3$&   4089  &   3575 &   3285 &   2682 \\
 & $z=0.5$&   3702 &   3142 &   2801 &   2713 \\
\hline
 $w$CDM & $z=0$ & 4923 & 4086 & 3746 & 3558  \\
 & $z=0.3$&   4118  &   3777 &   3765 &   3129 \\
 & $z=0.5$ &   3855  &   3855 &   2937 &   2927 \\
 \hline
\end{tabular}
\end{center} 
\end{minipage}
\end{table*}

\begin{table*}
\begin{minipage}{150mm}
\begin{center}
\caption{Average mass (in h$^{-1}$ M$_\odot$) of the most massive pair with relative velocity above three threshold values for the DEUS-FUR cosmologies at z = 0, 0.3 and 0.5 respectively. Haloes are detected with a linking length $b=0.15$ and the maximum separation for the pair is $d_{12}<10$ h$^{-1}$ Mpc.}
\label{table_multi2}
\begin{tabular}{clrrrrr}
\hline
 & v$_{12} \ge$ & 1500 km s$^{-1}$  & 2000 km s$^{-1}$  & 2500 km s$^{-1}$ & 3000 km s$^{-1}$ \\
\hline
RPCDM & $z=0$&   24.75 $\times 10^{14}$  &   21.07 $\times 10^{14}$  &    5.98 $\times 10^{14}$  &    1.97 $\times 10^{14}$   \\
 & $z=0.3$&   19.16 $\times 10^{14}$  &   19.16 $\times 10^{14}$  &    4.39 $\times 10^{14}$  &    1.31 $\times 10^{14}$   \\
 & $z=0.5$&   16.05 $\times 10^{14}$  &    5.10 $\times 10^{14}$  &    2.62 $\times 10^{14}$  &    2.26 $\times 10^{14}$ \\
\hline
$\Lambda$CDM-W7 & $z=0$&   49.40 $\times 10^{14}$  &   49.40 $\times 10^{14}$  &   49.40 $\times 10^{14}$  &   49.40 $\times 10^{14}$ \\
 & $z=0.3$&   28.06 $\times 10^{14}$  &   26.22 $\times 10^{14}$  &   23.96 $\times 10^{14}$  &    5.63 $\times 10^{14}$ \\
 & $z=0.5$&   22.41 $\times 10^{14}$  &   22.41 $\times 10^{14}$  &   16.78 $\times 10^{14}$  &    3.60 $\times 10^{14}$  \\
\hline
 $w$CDM & $z=0$&   49.40 $\times 10^{14}$  &   49.40 $\times 10^{14}$  &   49.40 $\times 10^{14}$  &   49.40 $\times 10^{14}$ \\
 & $z=0.3$&   35.76 $\times 10^{14}$  &   35.76 $\times 10^{14}$  &   33.17 $\times 10^{14}$  &   12.07 $\times 10^{14}$ \\
 & $z=0.5$ &   21.42 $\times 10^{14}$  &   21.42 $\times 10^{14}$  &   17.43 $\times 10^{14}$  &    4.65 $\times 10^{14}$ \\
 \hline
\end{tabular}
\end{center} 
\end{minipage}
\end{table*}

\end{document}